\documentclass[final]{aipproc}
\layoutstyle{6x9}

\usepackage{graphicx}
\usepackage{amsmath,amssymb}
\usepackage{lineno}

\newcommand{\rmd}{{\rm d}}
\newcommand{\rme}{{\rm e}}
\newcommand{\rmi}{{\rm i}}
\newcommand{\sgn}{{\rm sgn}}

\newcommand{\mgf}{{\cal B}}
\newcommand{\MGF}{\vec{\cal B}}
\newcommand{\AVP}{\vec{\cal A}}

\begin{document}


\title{Time-dependent approach to transport and scattering in atomic and mesoscopic physics}


\author{Tobias Kramer}{
  address={Institut f\"ur Theoretische Physik, Universit\"at Regensburg, 93040 Regensburg, Germany}
}

\begin{abstract}
Transport and scattering phenomena in open quantum-systems with a continuous energy spectrum are conveniently solved using the time-dependent Schr\"odinger equation. In the time-dependent picture, the evolution of an initially localized wave-packet reveals the eigenstates and eigenvalues of the system under consideration. We discuss applications of the wave-packet method in atomic, molecular, and mesoscopic systems and point out specific advantages of the time-dependent approach. In connection with the familiar initial value formulation of classical mechanics, an intuitive interpretation of transport emerges. For interacting many-particle systems, we discuss the efficient calculation of the self-consistent classical transport in the presence of a magnetic field.
\end{abstract}

\maketitle

\section{Introduction to time-dependent scattering theory}

Potential-scattering theory describes systems with a localized scattering region and a potential-free region far away from the scatterer. An incoming plane wave arrives at the scatterer and the changes of the phase and the direction of the incoming wave reveal properties of the scatterer. While this setup is convenient for the description of collision experiments with particle beams, in many other situations matter-waves originate from a localized region in space where a potential is present. The latter situation is the topic of this course. In this case the formulation and solution of scattering theory is best achieved in a time-dependent fashion. Instead of asymptotic potential-free regions in space, scattering is recast as an initial value problem, which provides a direct connection to the initial value problems familiar from classical mechanics. In quantum mechanics, the time-evolution of an initial state is governed by the time-dependent Schr\"odinger equation
\begin{equation}
[\rmi\hbar\partial_t-H]|\psi(t)\rangle=0.
\end{equation}
For propagating an arbitrary initial state, the basic object of interest is the propagator (or Feynman kernel) $K$, which advances the initial state $\psi(\mathbf{r},t')$ from initial time $t'$ to a later time $t$:
\begin{equation}
\psi(\mathbf{r},t)=\int_{-\infty}^\infty\rmd\mathbf{r}'\;K(\mathbf{r},t|\mathbf{r}',t')\psi(\mathbf{r}',t')
\end{equation}
For Hamiltonians which do not explicitly depend on time, many propagators are available in analytic form \cite{Grosche1998a}. In contrast, for explicit time-dependent Hamiltonians only few results are known, most of which are related to the existence of Ermakov invariants \cite{Ermakov2009,Dodonov1992a,Kramer2005a,Lohe2009}. If the initial state $\psi$ is an eigenstate of the Hamiltonian, the propagation will preserve the initial shape of the probability density and the state merely acquires a time-dependent phase, oscillating with the eigenenergy of the state. If we propagate any other state (which in principle can be decomposed into a superposition of eigenstates) the initial density distribution will change shape and move in space. A very instructive example is the Moshinsky shutter \cite{Moshinsky1952a}, where the sudden opening of the shutter gives rise to transient effects.
An important problem concerns the range of validity of the sudden and the adiabatic perturbation theories, which should be applicable for fast or slow changes of the Hamiltonian, respectively. For the case of the Moshinsky shutter with adjustable shutter opening-time the perturbative and the exact results are given and analyzed by Scheitler and Kleber \cite{Scheitler1988a}. Transient effects are also the topic of a recent comprehensive review article \cite{Campo2009a}. 
Besides transient effects, many systems display characteristic revival phenomena. Revivals bring back part of the time-evolved wave-packet to its initial position and can lead to complete or partial reconstructions of the initial density distribution \cite{Averbukh1989a}. Revival phenomena have been analyzed in detail for Rydberg states of the hydrogen atom, and also recently for electronic wave packets in mesoscopic systems \cite{Krueckl2009a,Romera2009}.
Here, we first give a brief introduction into the time- and energy-dependent propagator and Green function of the Schr{\"o}dinger equation, and proceed to discuss applications of the time-dependent picture in the photodetachment of negative ions. The next section shows how to solve the time-dependent Schr{\"odinger} equation numerically and how to obtain the spectrum of molecular systems. In the last section we model transport through semiconductor devices and finally discuss the effects on interactions and boundary conditions on the transport through nanodevices in a magnetic field.

\section{Wave-packet evolution and Green function}

The energy-dependent counterpart of the time-dependent kernel is the energy-dependent Green function
\begin{equation}\label{eq:resolvent}
G(\mathbf{r},\mathbf{r}';E)=\lim_{\eta\rightarrow 0_+}\langle \mathbf{r} | \frac{1}{E-H+\rmi\eta} | \mathbf{r}' \rangle
=\frac{1}{\rmi\hbar}\int_0^{\infty}\rmd t\;\rme^{\rmi (E+\rmi\eta) t/\hbar}K(\mathbf{r},t|\mathbf{r}',t'),
\end{equation}
where the limit $\eta\rightarrow 0_+$ selects the retarded solution. Only few closed forms of energy-dependent Green functions are known and the numerical integration over the highly oscillatory propagator fails to give converging results. These difficulties are related to the representation of the identity in position space as a $\delta$-distribution
\begin{equation}
K(\mathbf{r},0|\mathbf{r'},0)=\delta(\mathbf{r}-\mathbf{r}').
\end{equation}
The Fourier transform of the $\delta$-distribution in position representation is a constant function in momentum space and signifies that the energy-dependent Green function is the solution of the inhomogeneous stationary Schr{\"o}dinger equation for all energies
\begin{equation}\label{eq:green}
[E-H]G(\mathbf{r},\mathbf{r}';E)=\delta(\mathbf{r}-\mathbf{r}').
\end{equation}
However, in most physical applications the energy and the momenta are limited to a certain range of interest and thus it is possible to replace the $\delta$-distribution for example by a localized Gaussian function $S(\mathbf{r})$, which also represents a Gaussian momentum distribution. This step can also be done in a more formal way using coherent states, or by switching to the Bargmann representation \cite{Kramer2003b}. The imaginary part of the Green function encodes the spectrum of the system. This can be seen from the relation
\begin{equation}
\lim_{\eta\rightarrow 0}\left(\frac{1}{x+\rmi\eta}\right)=P\left(\frac{1}{x}\right)-\rmi\pi\sgn(\eta)\delta(x)
\end{equation}
Eq.~(\ref{eq:resolvent}) gives the local density of states at the location $\mathbf{r}\rightarrow\mathbf{r}'$
\begin{equation}
-\frac{1}{\pi}\Im[G(\mathbf{r},\mathbf{r};E)]=\langle\mathbf{r}|\delta(E-H)|\mathbf{r}\rangle=\int\rmd\lambda\sum_n \delta(E-E_{n,\lambda}){|\phi_{n,\lambda}(\mathbf{r})|}^2,
\end{equation}
where we introduced the complete set of eigenstates $\phi_{n,\lambda}$ with discrete ($n$) and continuous ($\lambda$) eigenvalues.
If we replace the $\delta$-distribution by a normalized state $S$, we obtain the weighted LDOS
\begin{equation}
-\frac{1}{\pi}\Im\left[\int\mathbf{r}'\int\mathbf{r}\;S(\mathbf{r}')^{*}G(\mathbf{r'},\mathbf{r};E)S(\mathbf{r})\right]=\langle S|\delta(E-H)|S\rangle,
\end{equation}
which represents the propagation of an initially localized wave-packet in the time domain, since
\begin{equation}
\Im[\langle S | \delta(E-H) | S \rangle]=\Im\left[\frac{1}{\rmi\hbar}\int_{0}^\infty\rmd t\;\rme^{\rmi E t/\hbar} \langle S(0)|S(t)\rangle\right],
\end{equation}
and
\begin{equation}
|S(t)\rangle=\rme^{-\rmi H t/\hbar}|S(0)\rangle.
\end{equation}
The autocorrelation function 
\begin{equation}
C(t)=\langle S(0) | S(t) \rangle
\end{equation}
is the cornerstone of the time-dependent approach to scattering theory. For a given Hamiltonian $H$ and initial state $|S(0)\rangle$, the autocorrelation function is obtained either numerically or analytically \cite{Heller1978a}. Depending on the physical system under consideration, the initial state $|S(0)\rangle$ has a direct physical interpretation \cite{Kramer2002a,Bracher2003a,Kramer2006b}, or alternatively is conveniently chosen as a vehicle to obtain the correlation function \cite{Kramer2008a}.
If the initial state is normailzed, the autocorrelation function starts with the value $C(0)=\langle S(0)|S(0)\rangle=1$. In an open system, where eventually all components of the wave-packet leave the initial region covered by $|S(0)\rangle$,  the autocorrelation function vanishes in the long-time limit $\lim_{t\rightarrow\infty}C(t)=0$. In a closed system this is not the case and the Fourier analysis of $C(t)$ reveals the discrete set of eigenenergies corresponding to the set of eigenstates in the system which are represented in the wave-packet.
For Hamilton operators which are maximally quadratic in positions and momenta, the propagator $K$ is given in closed form by Kolsrud \cite{Kolsrud1956a} based on unitary transformations. In this case the propagator is equivalent to a linear canonical transformation of positions and momenta. The position space representation of linear canonical transformations has been analyzed in detail by Moshinsky and Quesne \cite{Moshinsky1971a}. In all these cases, the quadratic form of the action preserves an initially Gaussian density-profile, and only the phase and width do evolve with time. The autocorrelation function and the spectrum of quadratic Hamiltonians cover a wide range of physical problems, some are discussed in the next sections.

\section{Photoionization spectra in external fields, closed orbit theory}

\begin{figure}[t]
\includegraphics[width=0.4\textwidth]{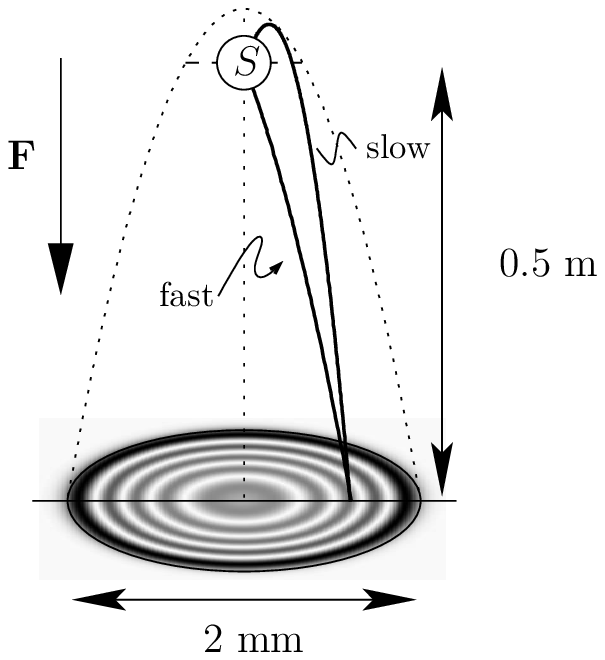}\hfill
\includegraphics[width=0.5\textwidth]{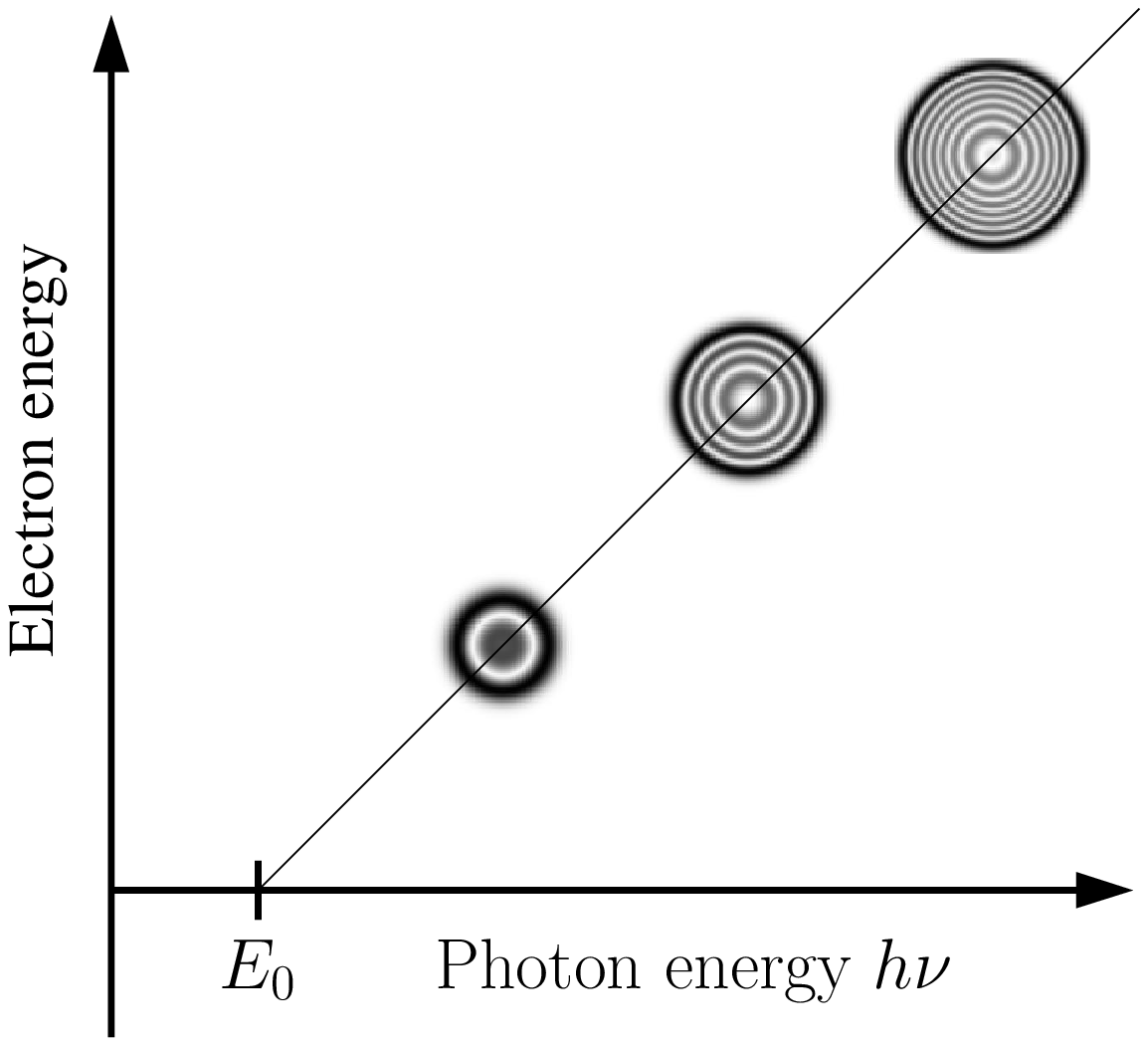}
\caption{\label{fig:photodet}Photodetachment microscopy in an electric field. The left panel shows the two parabolic pathways of the same energy linking the ion source $S$ and the detector. In the experiments by Blondel et.~al. the detector is placed $0.5$~m away from the ion source and the resulting electronic wave-function spreads out over several mm. The right panel depicts Einstein's law which relates the photon energy (known through the laser frequency) and the electron energy (determined by counting the number of interference rings). The intersection point at energy $E_0$ determines the electron affinity.}
\end{figure}

As a first application, we consider the photodetachment of an electron $e^{-}$ from a negative ion $X^-$ placed in external electric and magnetic. The ion is then radiated with a laser beam of photons $\gamma$ with energy $h\nu$. The detachment process is described by
\begin{equation}
X^{-}+\gamma \rightarrow X + e^{-}.
\end{equation}
Conservation of energy relates the photon energy $h\nu$, the electron energy $E(e^-)$, and the binding energy of the electron to the atom, the electron affinity $E_A(X^-)$
\begin{equation}
E_A(X^-)+h\nu=E(e^-),
\end{equation}
which is Einstein's law for the work-function applied to negative ions. If the wave-length of the emitted electron is large compared to the size of the emitting object (here the negative ion), it is possible to replace the emitter by a point source $S(\mathbf{r})=C\delta(\mathbf{r}-\mathbf{r}')$  and only consider the orbital characteristic explicitly \cite{Andersen2004,Bracher2003a}. We obtain an inhomogeneous Schr\"odinger equation for the electronic wave-function originating at $\mathbf{r}'$
\begin{equation}\label{eq:source}
[E-H]\psi(\mathbf{r};\mathbf{r}',E)=S(\mathbf{r}),
\end{equation}
where $C$ denotes the strength of the point source. The photodetachment process is modelled as a two-step process: first the energy gained by the photon-absorption promotes the electron from the bound state to an unbound state. In the second step, the residual effect of the remaining neutral atom $X$ is neglected and we consider the propagation in the potential given by the external fields, contained in the Hamiltonian $H$.

The measurement of the onset of the photodetachment current as function of photon-energy has been used to extract the electron affinity, but contains rather large experimental uncertainties. A much more precise method takes advantage of interference effects in order to measure the energy of the detached electron. Interference effects require the presence of external fields in order to construct multiple pathways from the ion to the electron detector, along which the electron travels coherently. The simplest configuration is the application of a uniform electric field in the Hamiltonian,
\begin{equation}\label{eq:Hfield}
H_{\rm field}=\frac{\mathbf{p}^2}{2m}-e\vec{{\cal E}}\cdot\mathbf{r}
\end{equation}
which creates a ``virtual double-slit''  (Fig.~\ref{fig:photodet}). This interferometric setup was proposed by Demkov, Kondratovich, and Ostrovskii \cite{Demkov1982a} and experimentally realized by Blondel, Delsart, and Dulieu \cite{Blondel1996a}. The observed quantity on the detector is the spatially-resolved rate of incoming electrons, the current density $\mathbf{j}(\mathbf{r},E)$ for different laser frequencies and thus for different electron energies $E$.

The solution of Eq.~(\ref{eq:source}) for a point-emitter is given by the Green function
\begin{equation}
\psi(\mathbf{r};E)=\int\rmd\mathbf{r}'\; G(\mathbf{r},\mathbf{r'};E) S(\mathbf{r}'),
\end{equation}
and the probability current-density becomes
\begin{equation}
\mathbf{j}(\mathbf{r};E)=\frac{\hbar}{m}\Im\left\{\psi(\mathbf{r};E)^*\nabla\mathbf{r};E)\right\}-\frac{e\AVP(\mathbf{r})}{m}{|\psi(\mathbf{r};E)|}^2,
\end{equation}
where $\AVP$ denotes the vector potential. Applying the equation of continuity 
\begin{equation}
\nabla\cdot \mathbf{j}(\mathbf{r};E)=-\frac{2}{\hbar}\Im\left\{S(\mathbf{r})^*\psi(\mathbf{r};E)\right\},
\end{equation}
allows us to obtain the total current by integrating over a surface enclosing the source $S(\mathbf{r})$
\begin{figure}[t]
\includegraphics[width=0.7\textwidth]{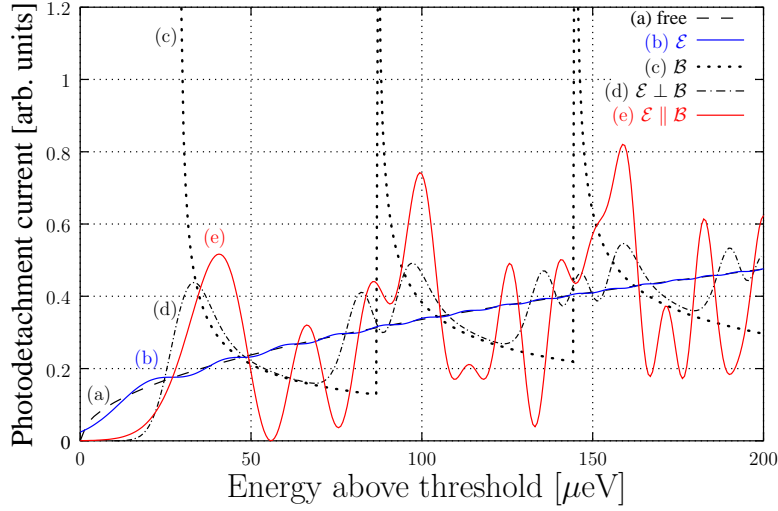}
\caption{Photodetachment currents for different electric and magnetic field configurations. The values of the magnetic field are ${\cal B}=0.5$~T and for the electric fields ${\cal E}=200$~V/m, the respective Hamilton operators are given in the text.\label{fig:photodetcomp}}
\end{figure}
\begin{equation}
J(E)=-\frac{2}{\hbar}   \Im\langle S | G | S \rangle .
\end{equation}
For a point-like source, the total current is proportional to the local density of states given by the imaginary part of the Green function
\begin{equation}
J(E)=-\frac{2}{\hbar} \Im G(\mathbf{r}',\mathbf{r}';E).
\end{equation}
For the field-free case we have
\begin{equation}\label{eq:Gfree}
G_{\rm free}(\mathbf{r},\mathbf{r}';E)=\frac{m}{2\pi\hbar^2|\mathbf{r}-\mathbf{r}'|}\exp\left[-|\mathbf{r}-\mathbf{r}'|\frac{\sqrt{-2mE}}{\hbar}\right],
\end{equation}
and the current near the detachment threshold has a square-root dependence on the laser energy
\begin{equation}\label{eq:Jfree}
J(h\nu)\propto \Theta(h\nu-|E_A|)\; \sqrt{h\nu-|E_A|}.
\end{equation}
The energy-dependent Green-function is known in closed form for very few problems \cite{Bracher2005b},  whereas the time-dependent propagator is available for a much larger class of problems \cite{Grosche1998a}. For all combinations of uniform electric and magnetic fields, a quadratic Hamilton operators results and therefore for initially Gaussian states the autocorrelation function can be obtained as analytical expression. Thus one can compute the spectrum at least numerically with very high precision for all field configurations. The resulting photodetachment currents are shown in Fig.~\ref{fig:photodetcomp} and display the strong influence of external fields on the photodetachment process, which causes large oscillations around the field-free current (a), Eq.~(\ref{eq:Jfree}). The presence of the electric field (eq.~\ref{eq:Hfield}) causes a stepwise modulation of the current (b), which is linked to the appearance of a new interference fringe in the spatially resolved current density \cite{Kramer2002a}. The case of parallel electric and magnetic fields (e) with Hamiltonian
\begin{equation}
H_{{\cal E}\parallel{\cal B}}=\frac{\mathbf{p}^2}{2m}-e {\cal E}z+\frac{1}{2}m\omega_l^2 (x^2+y^2)-\omega_l(y p_x-x p_y),
\quad \omega_l=\frac{e{\cal B}}{2m}
\end{equation}
is discussed in detail in Refs.~\cite{Kramer2001a,Bracher2005a}. For crossed electric and magnetic fields (d) with
\begin{equation}
H_{{E}\perp{\cal B}}=\frac{\mathbf{p}^2}{2m}-e {\cal E}x+\frac{1}{2}m\omega_l^2 (x^2+y^2)-\omega_l(y p_x-x p_y),
\quad \omega_l=\frac{e{\cal B}}{2m}
\end{equation}
an interesting substructure within magnetic Landau levels emerges, which is analyzed in Ref.~\cite{Kramer2003a}.

An elegant interpretation of the very prominent interference phenomena encoded in the total current is given by closed-orbit theory \cite{Peters1993a,Peters1993b,Peters1997b,Bracher2005a}. The basic idea is to approximate the Green function, given by the Laplace transform of the time-evolution operator
\begin{eqnarray*}
G(\mathbf{r},\mathbf{r}';E)
&=&-\frac{\rmi}{\hbar}\int_0^\infty\rmd t\,\rme^{\rmi E t/\hbar} 
\underbrace{\langle\mathbf{r}|\exp[-\rmi H t/\hbar]|\mathbf{r}'\rangle}_{\text{Feynman path integral}}\\
&=&-\frac{\rmi}{\hbar}\int_0^\infty\rmd t\,\rme^{\rmi E t/\hbar}\,a(\mathbf{r},\mathbf{r}',t) 
\exp[\rmi \underbrace{S(\mathbf{r},t|\mathbf{r}',0)}_{\text{action}}/\hbar],
\end{eqnarray*}
in the semiclassical limit $\hbar\rightarrow 0$. The saddle points are given by the following condition for the action
\begin{equation}
\frac{\partial}{\partial t} [E t+S(\mathbf{r},t|\mathbf{r}',0)]_{t=t_k}=0,
\end{equation}
which selects all classical trajectories corresponding to a fixed energy $E$. The travel times are denoted by $t_k$.
\begin{figure}[t]
\includegraphics[width=0.7\textwidth]{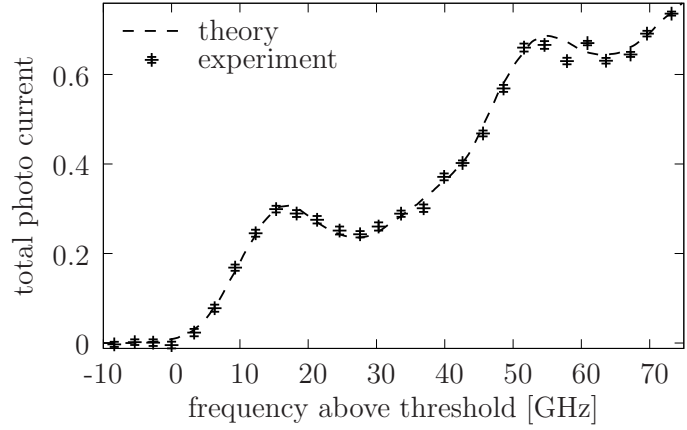}
\caption{Photodetachment current of negative sulfur in the presence of a magnetic field. The Zeeman splitting of the levels and the thermal occupation number are taken into account and require to superimpose the photodetachment currents of several levels (adapted from Ref.~\cite{Yukich2003a})\label{fig:photoexpthe}}
\end{figure}
The semiclassical Green function reads
\begin{equation}
G(\mathbf{r},\mathbf{r}';E)\approx
\underbrace{\overline{G}(\mathbf{r},\mathbf{r}';E)}_{\text{t=0}}+
\frac{1}{\rmi\hbar}\sum_{k=1}^N\!
a(\mathbf{r},\mathbf{r}',t_k)
\frac{\exp[\rmi[
E t_k+S(\mathbf{r},t_k|\mathbf{r}',0)]/\hbar]\rme^{\rmi\pi\text{sgn} [\ddot{S}(\mathbf{r},t_k|\mathbf{r}',0)]/4}}
{\sqrt{|\ddot{S}(\mathbf{r},t_k|\mathbf{r}',0)|/(2\pi\hbar)}},
\end{equation}
and represents a sum over classical paths weighted with complex phases. The first term represents the contribution of the $t=0$ pole of the propagator and requires a careful contour evaluation. The total current is related to all orbits returning to the point of origin and thus requires to identify the corresponding closed orbits in position space (not necessarily closed in momentum space). If besides the first derivative of the action also the second derivative vanishes, the primitive semiclassical approximation diverges (diffraction catastrophe) and higher order terms are required. The challenging evaluation of the corresponding diffraction integrals is a topic of catastrophe theory \cite{Bracher2005a,Poston1978a}. Measurement in parallel and perpendicular fields by Yukich \cite{Yukich2003a} and Blondel et~al \cite{Blondel2010a} are in excellent agreement with the theoretical description given above. In Fig.~\ref{fig:photoexpthe} we compare the theoretically calculated photodetachment current of the sulfur-ion in the presence of crossed electric and magnetic fields (${\cal B}\approx$~1 Tesla) with the experimental data \cite{Yukich2003a}.

For coherent quantum sources, which have an extension of the order of the wavelength of the emitted particles, the source structure does influence the cross-section and has to be taken into account. An example of such a system is the atom-laser from macroscopic Bose-Einstein condensates, where a weak perturbation by a radio-field causes a coherent outcoupling of atoms from a trapped condensate \cite{Bloch1999a,Bloch2000a}. Depending on the size of the BEC interference phenomena will occur or a tunneling regime prevails with only one possible trajectory \cite{Kramer2002a,Bracher2003a,Kramer2006b}.

\section{Molecular physics}

\begin{figure}[t]
\includegraphics[width=0.45\textwidth]{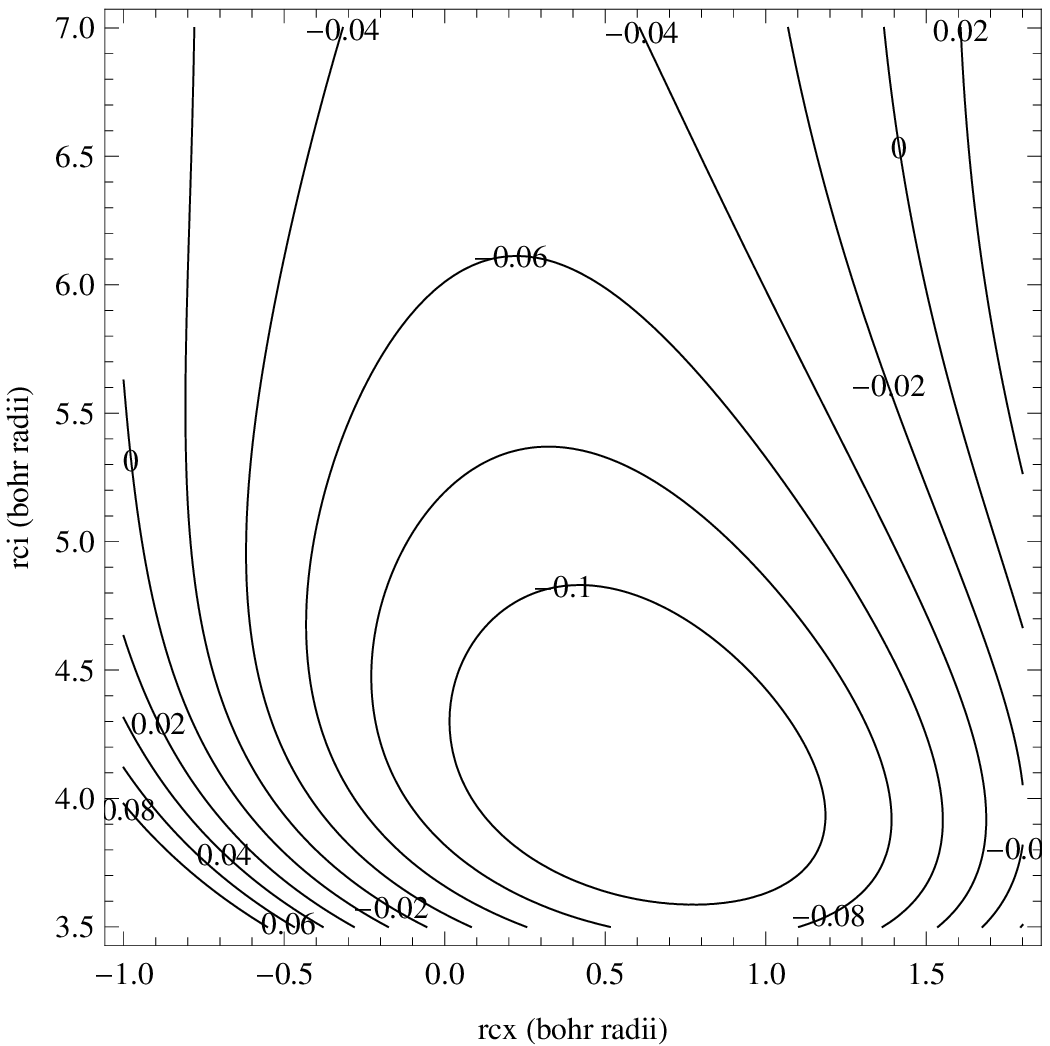}\hfill
\includegraphics[width=0.45\textwidth]{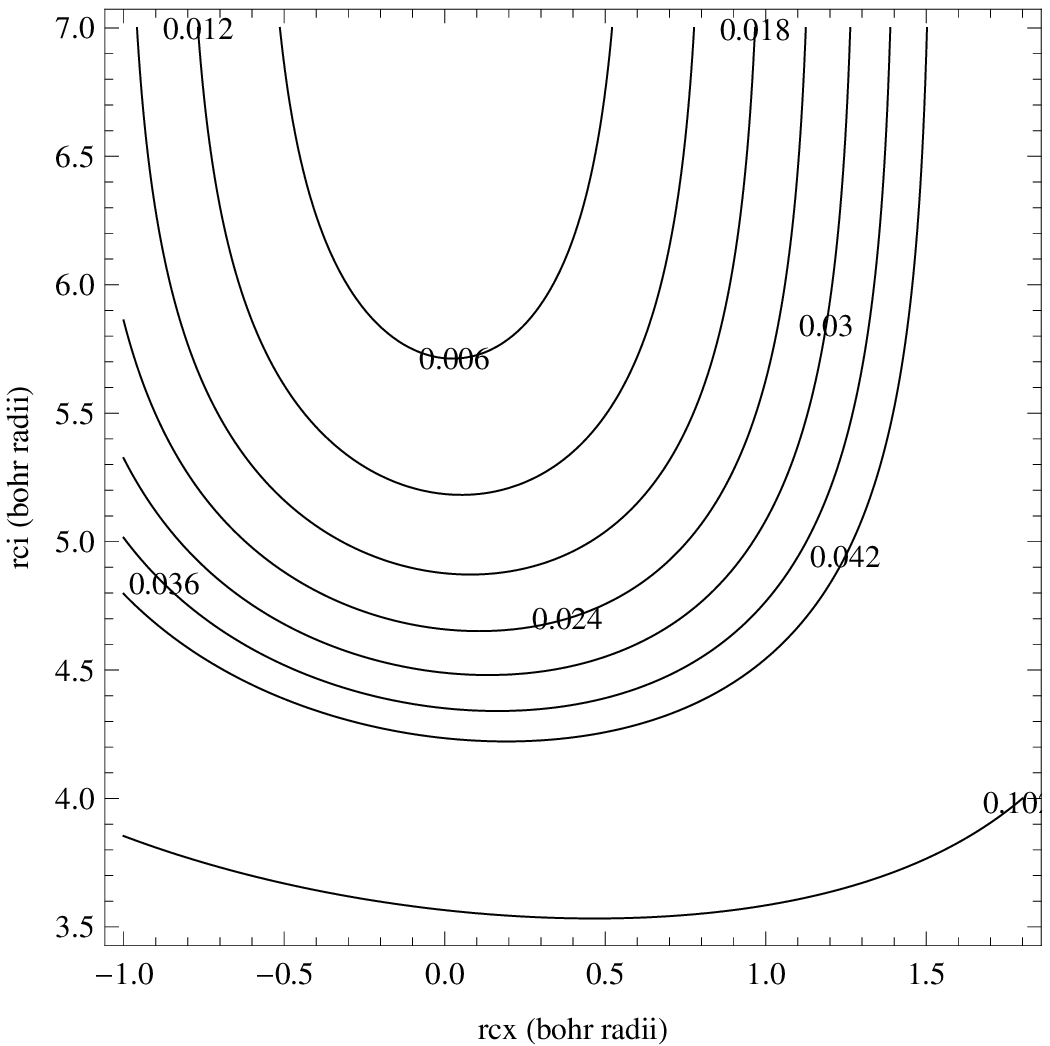}\\
\caption{Contour plots of the potential energy surface of $CH_3I$ for the ground state (left panel) and the excited state (right panel). Countour labels denote the energies in Hartrees.
}\label{fig:pes_ch3i}
\end{figure}

The time-dependent approach to molecular physics was explored in the seventies by Heller \cite{Heller1978a,Heller1981} and is the topic of a recent textbook by Tannor \cite{Tannor2007a}. Wave-packets are extremely valuable tools in theoretical chemistry and have been used to describe photochemical reactions, such as the photodissociation of molecules \cite{Lee1982a}. In quantum-chemistry, the potentials are generally not available in analytic form and numerical methods are required to propagate any initial state. Moreover, often the molecular kinetic energy operator is coupled to spatial coordinates which further complicates the analysis.

As a specific example we discuss the photodissociation of a linear, triatomic molecule following the time-dependent approach of Ref.~\cite{Lee1982a}. We describe the molecular Hamiltonian within the Born-Oppenheimer approximation, which yields the potential energy surface (PES) for the motion of the nuclei, given by the addition of the nuclear Coulomb energies and the electronic energy of a specific electronic state for a fixed nuclear configuration. The electrons are assumed to follow the nuclear motion instantaneously and cross-couplings between the electronic and nuclear momenta are neglected. The introduction of Jacobi coordinates facilitates the discussion and separates the center-of-mass motion from the relative motion. To be specific, let as consider methyl-iodine $CH_3I$, which has a tetrahedral structure with the 3 hydrogen atoms arranged in a plane, the carbon atom situated above with the iodine attached and pointing outwards normal to the hydrogen plane. For simplicity, we will collapse the three hydrogen atoms to a single ``atom'' (denoted by $X=3H$). The kinetic energy of the system is then given by the sum of the kinetic energies of the three nuclei
\begin{equation}
T=\frac{p_I^2}{2M_I}+\frac{p_C^2}{2M_C}+\frac{p_X^2}{2M_X}.
\end{equation}
Jacobi coordinates for $N$ particles located at positions $\mathbf{r}_1,\ldots,\mathbf{r}_N$ are specified by
\begin{eqnarray}
\vec{\xi}_1&=&\frac{M_1\mathbf{r}_1}{M_1}-\mathbf{r_2}=\mathbf{r}_1-\mathbf{r}_2,\\
\vec{\xi}_2&=&\frac{M_1\mathbf{r}_1+M_2\mathbf{r}_2}{M_1+M_2}-\mathbf{r_3},\\
\vec{\xi}_j&=&\frac{M_1\mathbf{r}_1+\cdots+M_j\mathbf{r}_j}{M_1+\cdots+M_j}-\mathbf{r_{j+1}},\\
\vec{\xi}_N&=&\frac{M_1\mathbf{r}_1+\cdots+M_N\mathbf{r}_N}{M_1+\cdots+M_N}=\mathbf{R}.
\end{eqnarray}
\begin{figure}[t]
\includegraphics[width=0.49\textwidth]{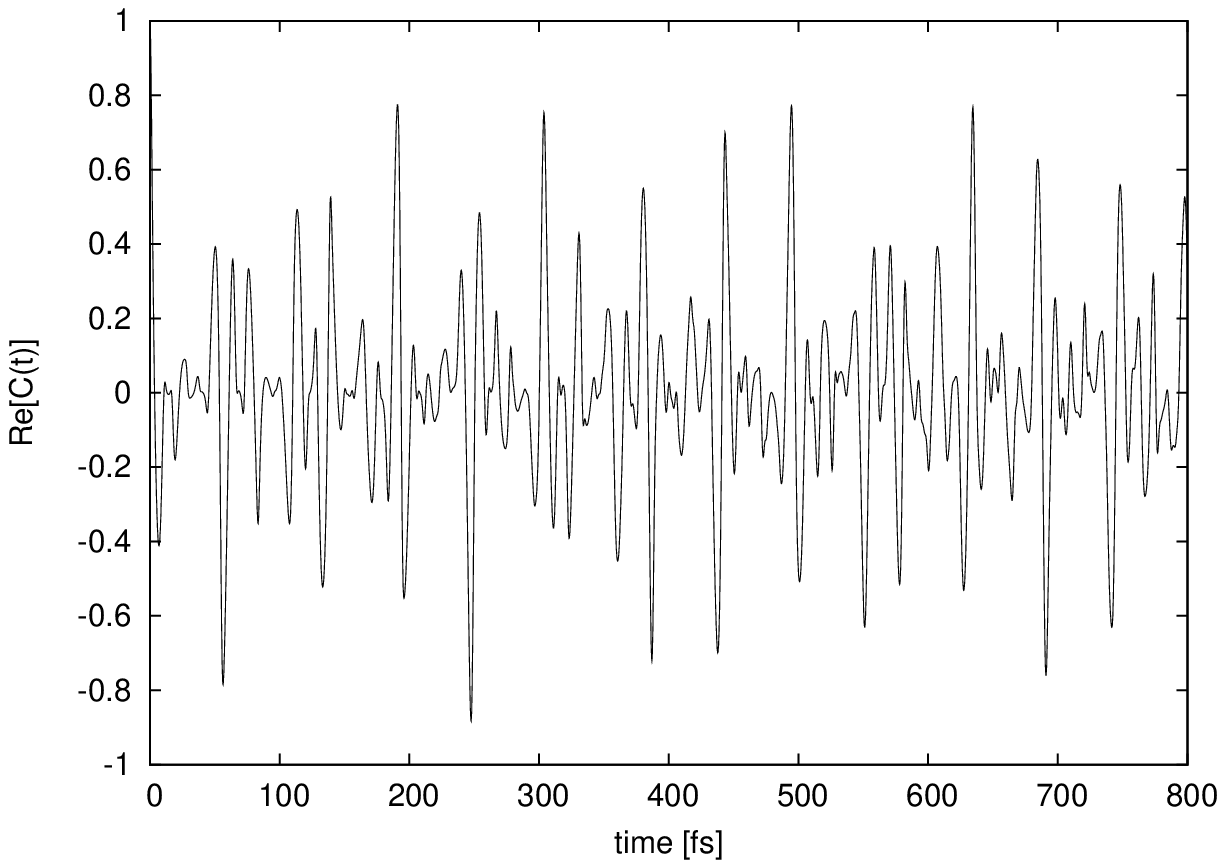}\hfill
\includegraphics[width=0.49\textwidth]{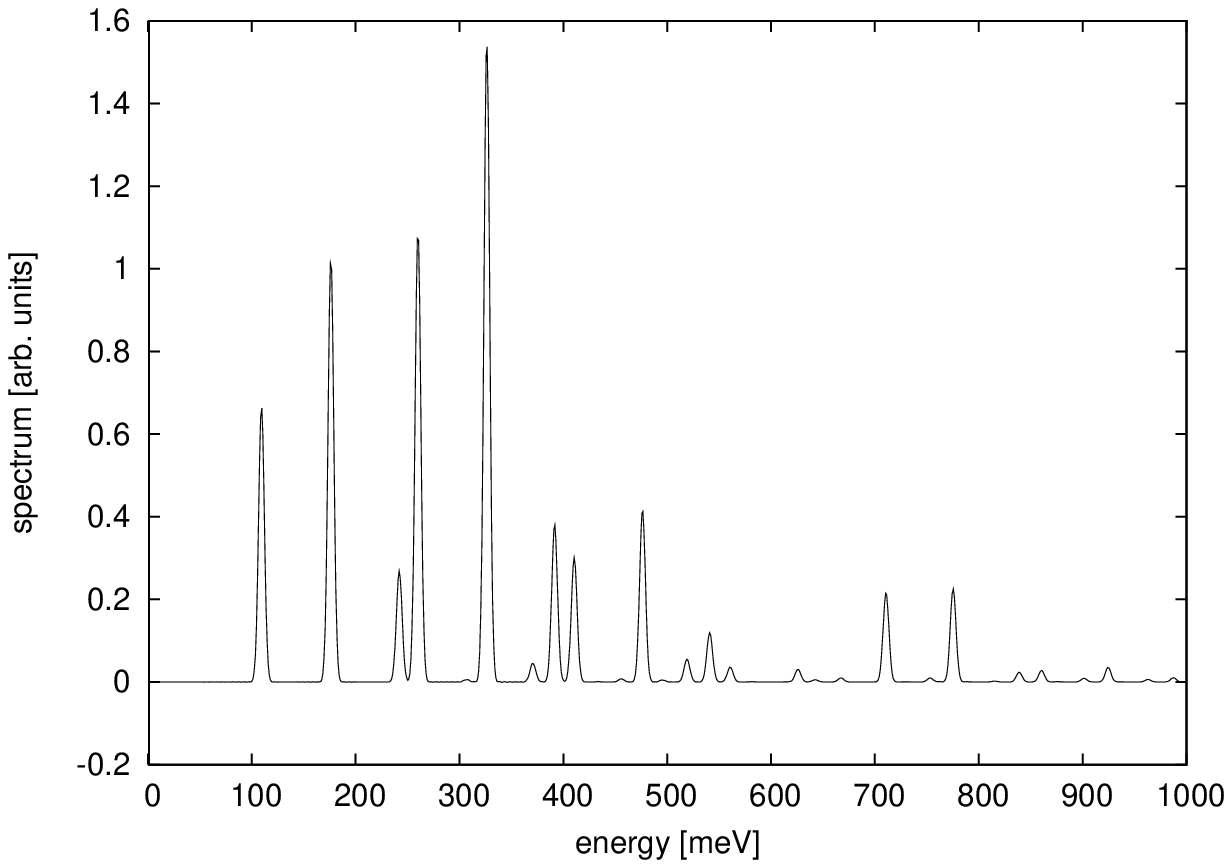}
\caption{Autocorrelation function (left panel) and its Fourier-transform, which reveals eigenstates with eigenenergies at $E_0=109$~meV, $E_1=176$~meV, $E_2=242$~meV, $E_3=260$~meV, $E_4=326$~meV.\label{fig:ch3i_act}}
\end{figure}
Together with the total mass and the reduced mass
\begin{equation}
M=\sum_{i=1}^N M_i,\quad
\mu_i^{-1}={\bigg(\sum_{j=1}^i M_j\bigg)}^{-1}+M_{i+1}^{-1},
\end{equation}
the Hamiltonian is expressed as
\begin{equation}
H=H_{\rm cm}+H_{\rm rel}=-\frac{\hbar^2}{2M}\nabla_{\mathbf{R}}^2+\sum_{i=1}^{N-1}-\frac{\hbar^2}{2\mu_i}\nabla_{\xi_i}^2+V(\vec{\xi}_1,\ldots,\vec{\xi}_{N-1}).
\end{equation}
If we consider $CH_3I$ as a linear triatomic molecule, where the postions of the three atoms are described by the three coordinates $r_I,r_C,r_X$, the Jacobi coordinates reduce the problem to a two-dimensional one, where only the distances  $\xi_1=r_C-r_X$ and $\xi_2=r_{\rm cm}(CX)-r_I$ appear. For the atomic weights $M(X)=M_1=3u$, $M(C)=M_2=12u$, $M(I)=M_3=127u$, we obtain $\xi_1=r_{CX}$ and $\xi_2=r_{CI}+\frac{3}{15}r_{CX}$, $\mu_1=\frac{3\cdot 12}{3+12} u$, $\mu_2=\frac{15\cdot 127}{15+127}u$.
For the potential $V(\xi_1,\xi_2)$ we use the expressions given in Ref.~\cite{Shapiro1980}
\begin{eqnarray}
V_{\rm gr}(r_{CX},r_{CI})&=&-D_e-E^*
+D_e{[\rme^{-\beta (r_{CI}-r_{CI}^e)}-1]}^2\notag\\
&&+\frac{1}{2}[k+(k^e-k)\rme^{-\alpha(r_{CI}-r_{CI}^e)}]
{[r_{CX}-r_{CX}^e \rme^{-\alpha(r_{CI}-r_{CI}^e)} ]}^2,
\end{eqnarray}
\begin{figure}[t]
\begin{minipage}{\textwidth}
\begin{center}
\includegraphics[width=0.325\textwidth]{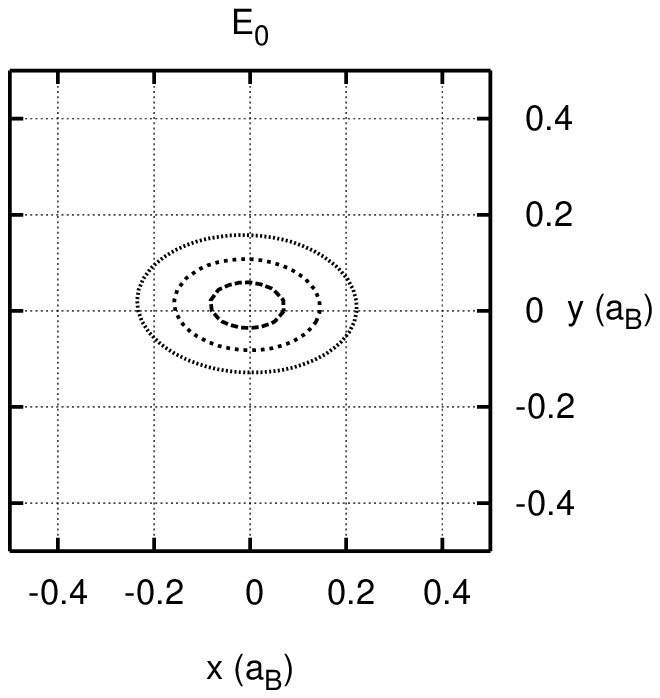}
\includegraphics[width=0.325\textwidth]{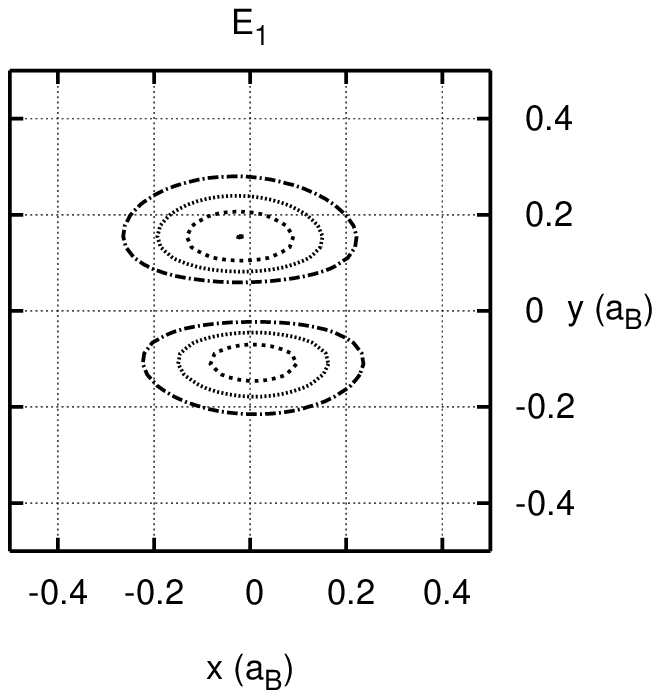}
\includegraphics[width=0.325\textwidth]{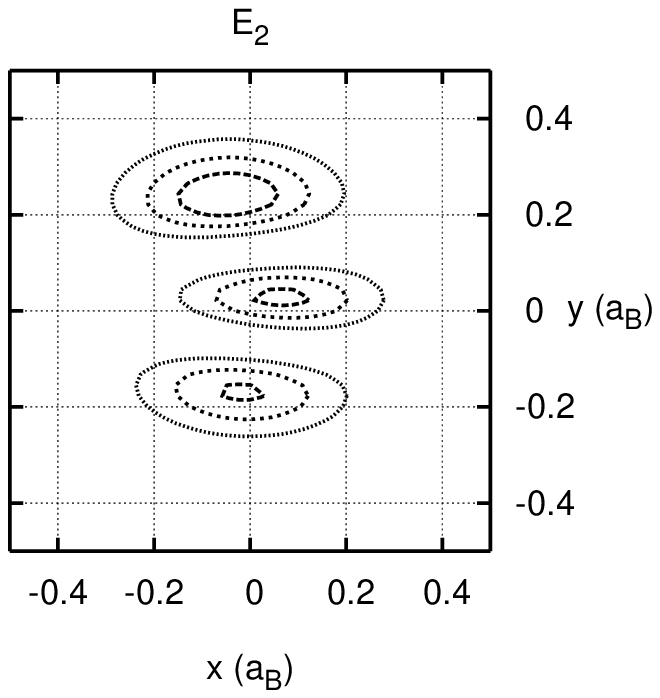}\\
\includegraphics[width=0.325\textwidth]{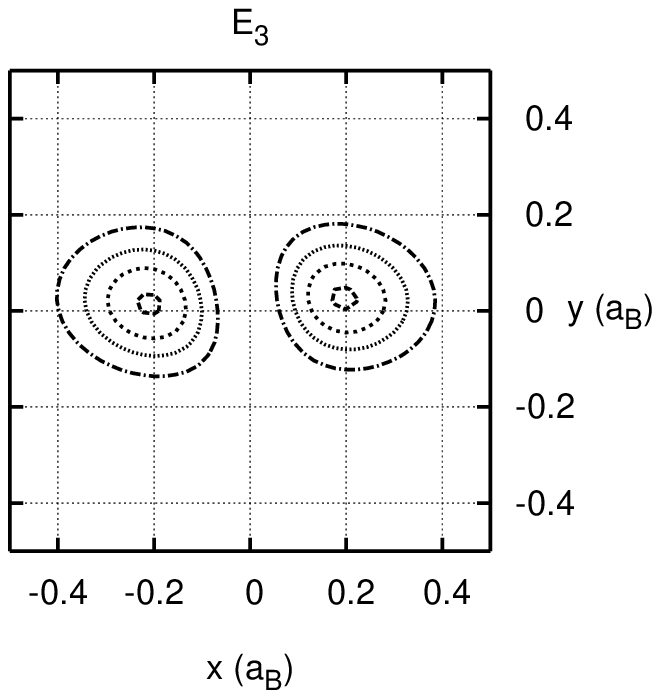}
\includegraphics[width=0.325\textwidth]{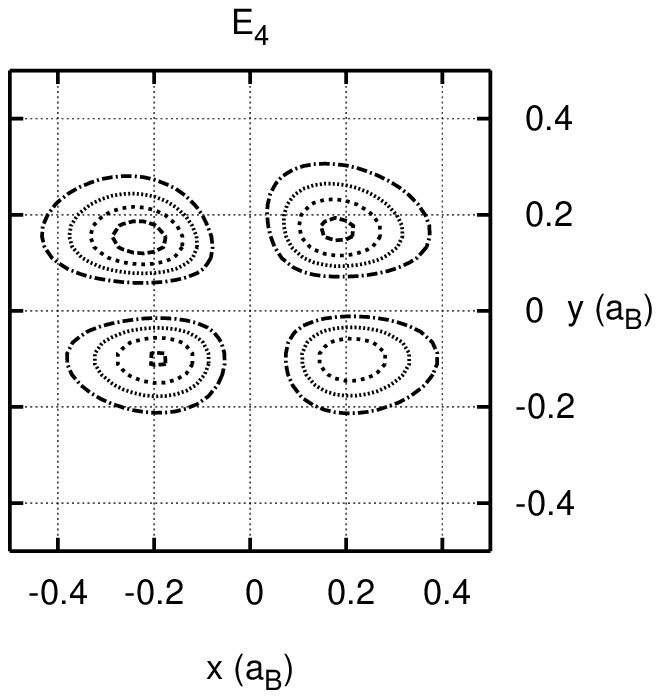}
\end{center}
\end{minipage}
\caption{Probability densities of the lowest five energy-eigenstates corresponding to the labelled peaks in the spectrum in Fig.~\ref{fig:ch3i_act}, obtained by propagating an Gaussian wavepacket and projecting out eigenstates with Eq.~(\ref{eq:eigenproj}). \label{fig:ch3ieigenstate}}
\end{figure}
with all length measured in Bohr radii $a_B=5.29\cdot 10^{-11}$~m and the parameters $D_e=0.0874$~Hartree, $E^{*}=0.0346$~Hartree, $\alpha=0.4914/a_B$, $\beta=0.899/a_B$, $r_{CI}^e=4.043\;a_B$, $r_{CX}^e=0.6197\;a_B$, $k=0.0363$~Hartree$/a_B^2$, $k^e=0.1463$~Hartree$/a_B^2$.
The potential energy surface is shown in Fig.~\ref{fig:pes_ch3i}. In the first step we propagate an arbitrary wavepacket (for example a Gaussian of width $a=5\cdot10^{-12}$~m) over 4000 timesteps $\Delta t=2\cdot10^{-16}$~s on a grid of $[-1.5,+1.5]\times[-1.5,+1.5]$~${(10^{-10}\text{m})}^2$. Two algorithms are commonly used, the split-operator method based on Trotter's formula \cite{Feit1982a,Lubich2008} and the direct polynomial expansions of the time-evolution operator $\rme^{-\rmi H t/\hbar}$ \cite{Kosloff1994a}. The split-operator method relies on the Fast Fourier Transfrom (FFT) method to apply momentum and position-dependent operators by simple multiplications to the wavefunction in position and momentum representation. The symmetrized split-operator algorithm is accurate up to order ${\Delta t}^3$:
\begin{eqnarray}\nonumber
\psi(\mathbf{r},t'+N\Delta t)&\approx&
\rme^{-\rmi\Delta t/\hbar\;V(\mathbf{r})/2}\\
&&\times
{\left[
\rme^{-\rmi\Delta t/\hbar\;V(\mathbf{r})}
{\cal F}^{-1}
\rme^{-\rmi\Delta t/\hbar\;T(\mathbf{p})}
{\cal F}
\rme^{-\rmi\Delta t/\hbar\;V(\mathbf{r})}
\right]}^N\nonumber\\
&&\quad\times\rme^{\rmi\Delta t/\hbar\;V(\mathbf{r})/2}\psi(\mathbf{r},t').
\end{eqnarray}
Here, ${\cal F}$ denotes the Fourier transforms and ${\cal F}^{-1}$ the inverse Fourier transform. The resulting autocorrelation function is displayed in the left panel of Fig.~\ref{fig:ch3i_act}. The peaks of the Fourier transform of $C(t)$ indicate the postion of eigenfunctions (right panel in Fig.~\ref{fig:ch3i_act}). At these energies, we project out in a second run the eigenstates by recording
\begin{equation}\label{eq:eigenproj}
\psi_{E_i}(\vec{\xi})=\int_0^{T}\rmd t\;\rme^{-\rmi E_i t/\hbar}\psi(\vec{\xi},t).
\end{equation}
Fig.~\ref{fig:ch3ieigenstate} shows the lowest five energy eigenstates.
\begin{figure}[t]
\includegraphics[width=0.49\textwidth]{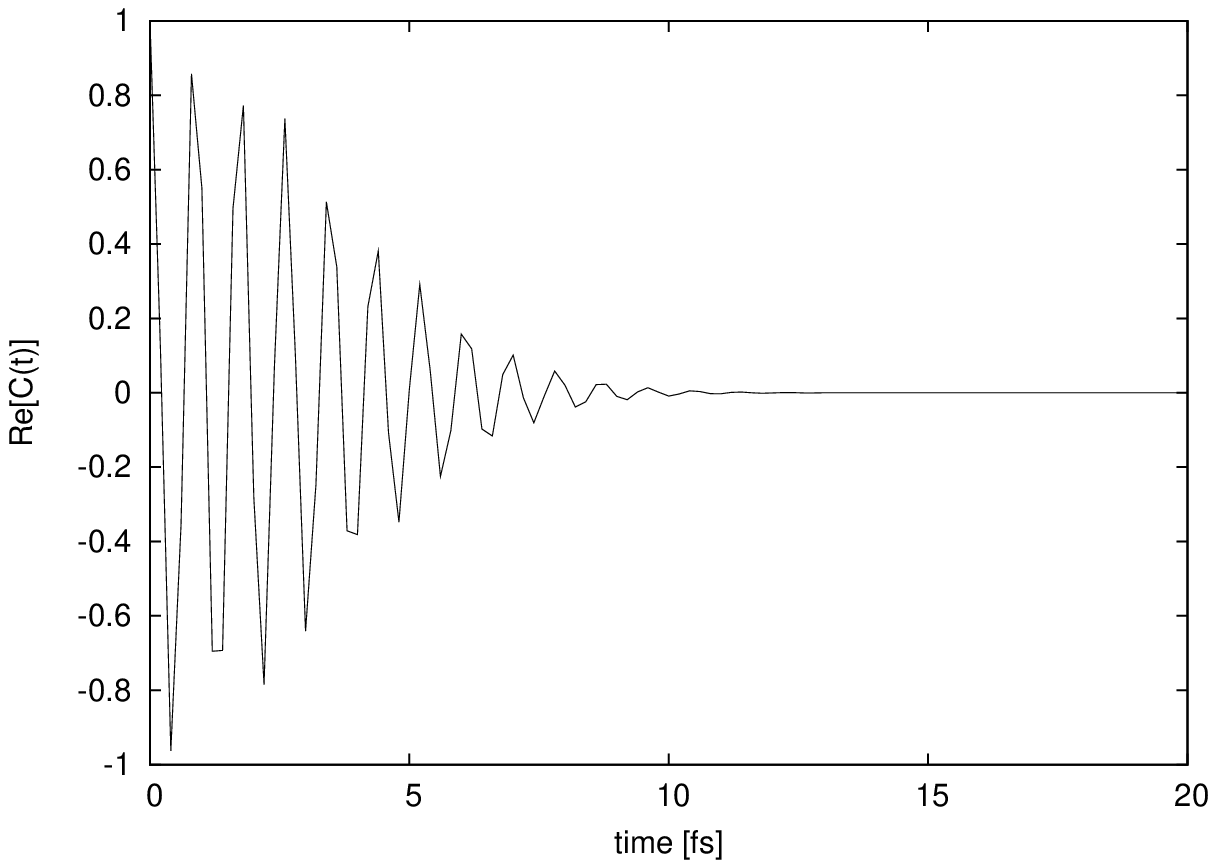}\hfill
\includegraphics[width=0.49\textwidth]{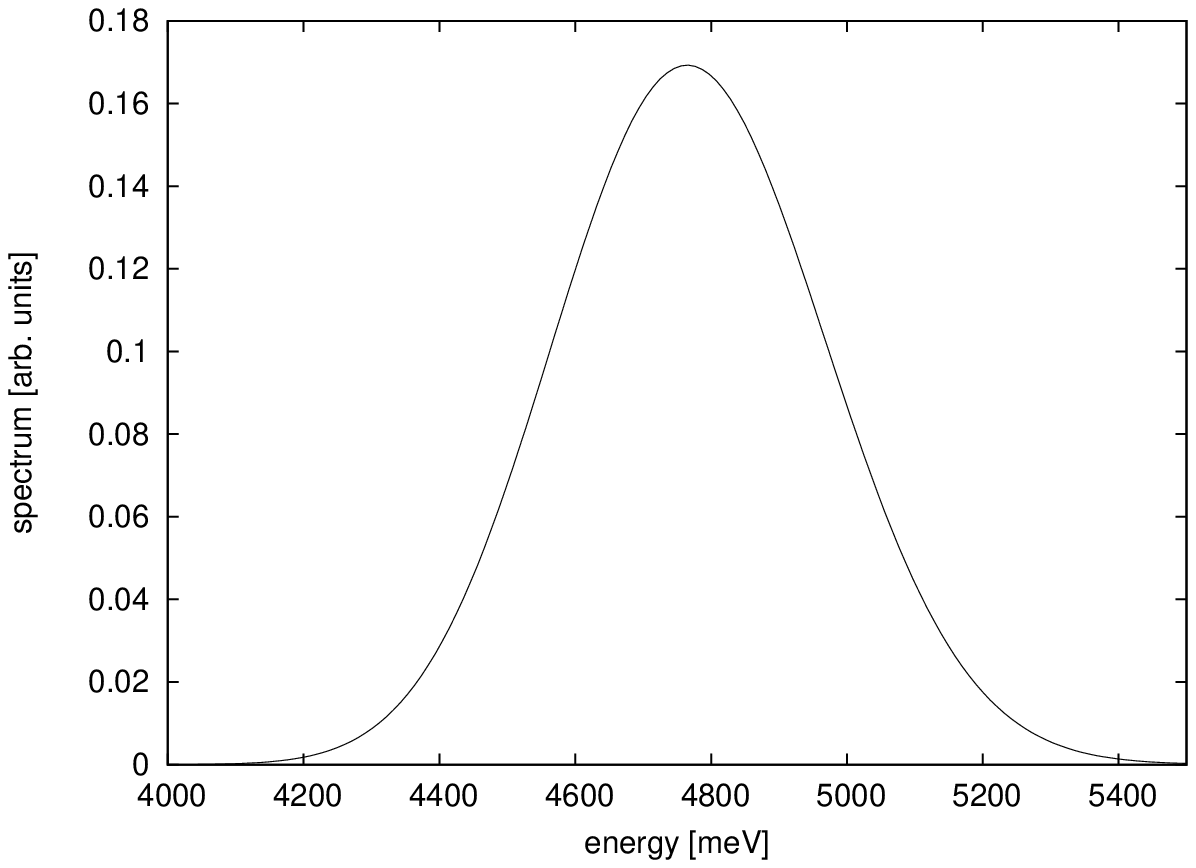}
\caption{Autocorrelation function (left panel) and its Fourier-transform resulting from evolving the ground-state $\psi_{E_0}(\vec{\xi})$ on the excited state potential energy surface. The continuous spectrum is directly proportional to the photodissociation cross-section.\label{fig:ch3i_ex_act}}
\end{figure}
To describe a photoionization process, we need the potential energy surface of the excited molecule, where the iodine is in a new electronic state
\begin{equation}
CH_3I+\gamma\rightarrow CH_3+I^{*}({}^2P_{1/2}),
\end{equation}
separated from the ground state energy by the energy of the photon. The PES of the excited state from Ref.~\cite{Shapiro1980} reads in Hartrees
\begin{equation}
V_{\rm ex}(r_{CX},r_{CI})=9.618\rme^{-1.4 (r_{CI}+0.2 r_{CX})/a_B}+2.604\rme^{-1.2 r_{CI}/a_B}+\frac{1}{2} 0.0362 \frac{r_{CX}^2}{a_B^2},
\end{equation}
and is shown in Fig.~\ref{fig:pes_ch3i}. In addition, also the knowledge of the dipole moments $\mu_d$ is required, here we choose for simplicity $\mu_d=1$. The cross-section of the photoreaction is given by propagating the ground state $\psi_{E_0}$ on the excited potential energy surface,
\begin{equation}
S(\vec{\xi};t)=\rme^{-\rmi t (T_{\rm rel}+V_{\rm ex})/\hbar}\left[\mu_d \psi_{E_0}(\vec{\xi})\right]
\end{equation}
recording the autocorrelation function 
\begin{equation}
C(t)=\int\rmd\vec{\xi}\;S^*(\vec{\xi};0)S(\vec{\xi};t),
\end{equation}
and finally applying the Fourier transform to $C(t)$, which results in the spectrum displayed in Fig.~\ref{fig:ch3i_ex_act}. The excited PES does not support bound states and thus the autocorrelation function decays with time $\lim_{t\rightarrow\infty}C(t)=0$. This decay implies a continuous spectrum for the process under consideration. The molecule $CH_3I$ still serves as a prototype for calculating PES and propagating wave-packets and recent results are given in Refs.~\cite{Nalda2008a,Alekseyev2007b}.

We have described the complete reaction dynamics using only time-dependent methods and without need to diagonalize matrices. The last point is important for the application of the time-dependent theory to transport in mesoscopic systems.

\section{Transport through mesoscopic systems}

\begin{figure}[t]
\includegraphics[width=0.7\textwidth]{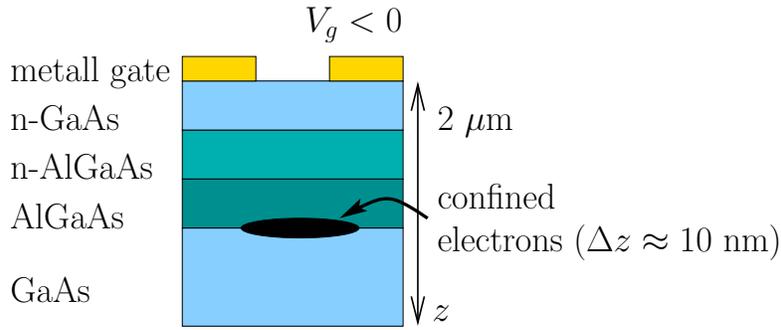}
\caption{AlGaAs/GaAs heterostructure with different layers of semiconductors. The quantum well forms at the interface between AlGasAs and GaAs and is filled with electrons donated from silicon atoms.
}\label{fig:nanostruct}
\end{figure}

In this section we consider many-electron systems in semiconductors at low temperatures. An introduction to the physics of mesoscopic systems present in nanostructures is given by Davies and Ferry \cite{Davies1998a,Ferry1997a}. AlGaAs/GaAs heterostructures are nanodevices consisting of layers of different semiconductor crystals, which have slightly different band structures. At the two-dimensional interface between two materials, the mismatch in the conduction- and valence-band results in the formation of a potential well, which traps and confines electrons along the two-dimensional interface (see Fig.~\ref{fig:nanostruct}). The electrons in the well originate from implanted donor atoms (for example silicon atoms), which are spatially separated from the interface. This separation in combination with extremely clean crystal growth using molecular beam epitaxy (MBE) leads to a high electron mobility and the suppression of scattering events in the well. At liquid helium temperatures (4~K), the electrons can transverse distances of several $\mu$m without loss of coherence.

The strong confinement of the mobile electrons at the interface reduces the intially three-dimensional problem to an effectively two-dimensional one, which is very well suited for using wave-packet techniques. Refs.~\cite{Kramer2008a,Kramer2010a} contain a detailed description of the time-dependent approach to transport in mesoscopic systems. Again, the key quantity is the time-dependent correlation function. In experiments, magnetic fields are commonly used to alter the electronic pathways and to exploit additional magnetic-flux dependent phase effects. In the presence of a magnetic field perpendicular to the interface, the split-operator method has to be modified since the Hamiltonian contains products of the momentum and position operator which require to divide the Fourier-transform step into two parts. The vector potential for a homogeneous magnetic field $\MGF=(0,0,{\cal B})$ in the symmetric gauge becomes $\AVP=(-y,x,0) {\cal B}/2$. The Hamiltonian splits into three parts
\begin{equation}
H=\underbrace{{\frac{p_x^2}{2m}-\omega_L p_x y}}_{T_{p_x,y}}
+\underbrace{\frac{p_y^2}{2m}+\omega_L p_y x}_{T_{p_y,x}}
+\underbrace{\frac{1}{2}\omega_L^2 (x^2+y^2)+V(x,y)}_{V_{\rm eff}(x,y)},\quad
\omega_L=\frac{e {\cal B}}{2m},
\end{equation}
where the mixed momentum-position representation for the kinetic energy is possible since $[p_x,y]=[p_y,x]=0$. The new propagation algorithm reads
\begin{eqnarray}\nonumber
\psi(\mathbf{r},t'+N\Delta t)&\approx&
\rme^{-\rmi\Delta t/\hbar\;V_{\rm eff}/2}\\
&&\times
{\left[
\rme^{-\rmi\Delta t/\hbar\;V_{\rm abs}}
{\cal F}^{-1}_y
\rme^{-\rmi\Delta t/\hbar\;T_{py,x}}
{\cal F}_y
{\cal F}^{-1}_x
\rme^{-\rmi\Delta t/\hbar\;T_{px,y}}
{\cal F}_x
\rme^{-\rmi\Delta t/\hbar\;V_{\rm eff}}
\right]}^N\nonumber\\
&&\quad\times\rme^{\rmi\Delta t/\hbar\;V_{\rm eff}/2}\psi(\mathbf{r},t'),
\end{eqnarray}
where ${\cal F}_x, {\cal F}_y$ denote partial Fourier transforms with respect to only one-dimension.

\begin{figure}[t]
\includegraphics[width=0.43\textwidth]{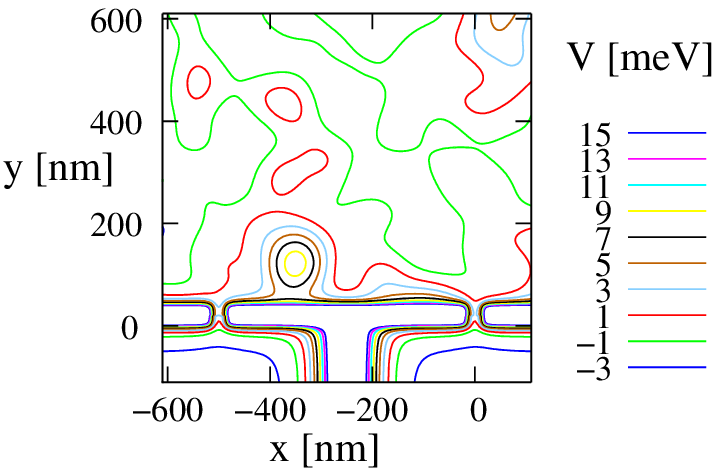}\hfill
\includegraphics[width=0.53\textwidth]{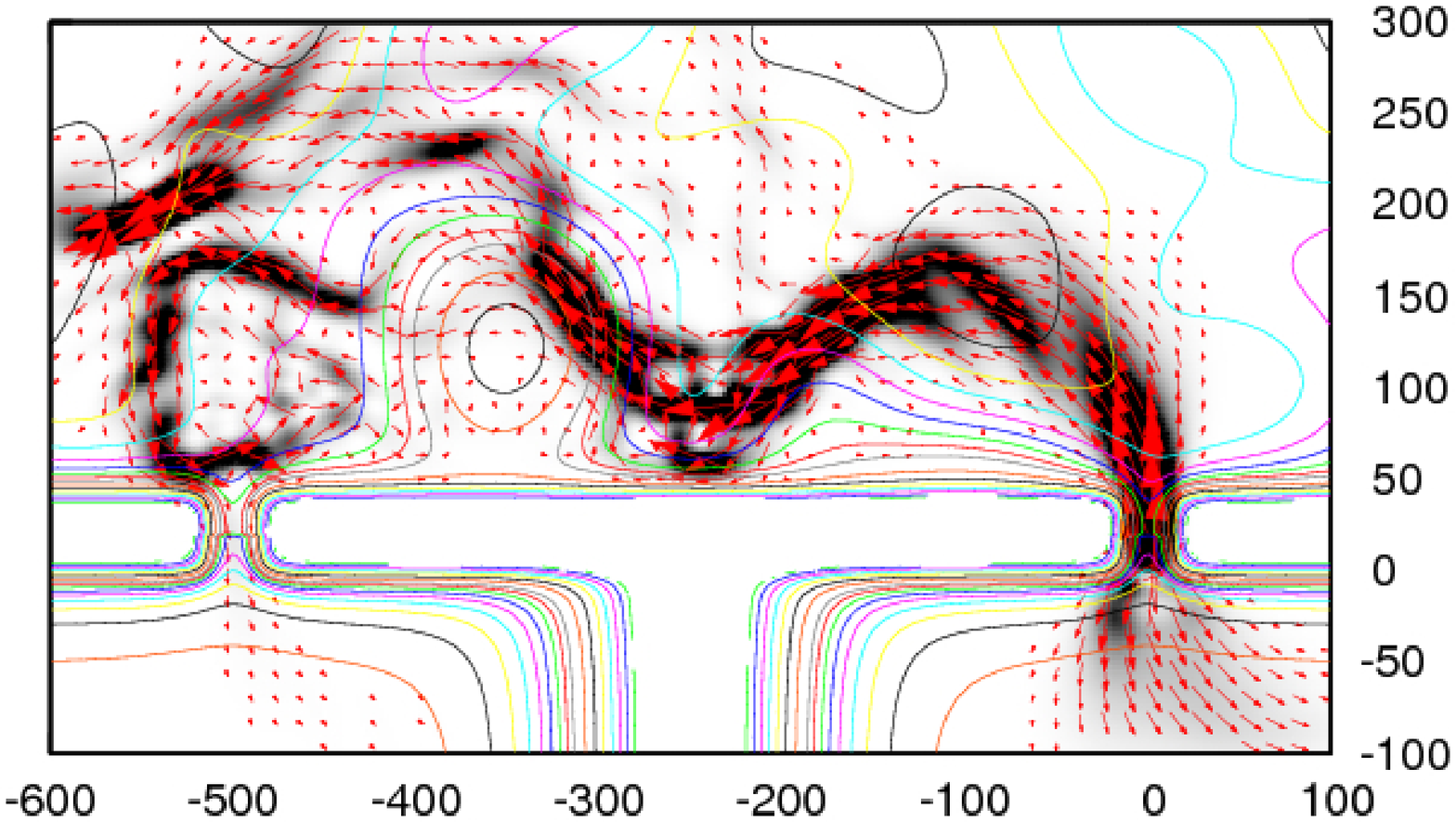}
\caption{Left panel: two-dimensional potential landscape of the device. The T-shaped potential of two quantum point contacts (at $x=-500$~nm and $x=0$~nm) forms two constrictions, while the background charges produce an irregular potential landscape. Right panel: gray-scale representation of the probability density and flux (arrows), obtained by starting a wave-packet located at $(x,y)=(0,0)$ and Fourier-analyzing the time-evolution at the energy $E=10$~meV. The obstacle at $(-350,125)$~nm interrupts the electron flow from the right chamber to the left one and results in a reduction of the transmission.
}\label{fig:mf_pot}
\end{figure}

\subsection{Magnetic focussing}

The effects of a magnetic field on electron transport are particularly strong in the magnetic-focussing configuration shown in Fig.~\ref{fig:mf_pot}. The electric current originates from the lower left chamber and has to pass through a Quantum Point Contact (QPC), which is in fact not a pointwise object, but rather a narrow constriction. Afterwards the electrons are following a circular trajectory, whose radius is proportional to the inverse of the magnetic field
\begin{equation}
r_c=\frac{\sqrt{2 m E}}{e{\cal B}}.
\end{equation}
If multiples of the radius $r_c$ equal the spacing to another constriction located at $x=-500$~nm, current can enter the left chamber. The macroscopic current flow from the right chamber to the left one is therefore strongly magnetic field dependent. So far we have neglected any additional potential perturbation caused from residual donor charges. In a real device, these charges lead to potential fluctuations. The fluctuations are on a scale of 10 percent of the Fermi-energy $E_F$ of the electron system (typical values of the Fermi energy are $E_F=10$~meV). The effective potential $V(x,y)$ of the device is shown in the left panel of Fig.~\ref{fig:mf_pot}. The transport paths at the Fermi-energy through the device are revealed by propagating a wave-packet starting in the lower right chamber and tracking its time-evolution. By Fourier-transforming the time-evolved wave-function, we obtain the stationary state connecting both chambers at the Fermi-energy of the electron gas, visible in the right panel of Fig.~\ref{fig:mf_pot}.

A direct imaging and verification of the theoretically calculated transport-pathways is possible using scanning-probe microscopy. In these experiments, the electronic pathways are locally perturbed by inducing a bump in the potential with a metallic tip, placed closely above the device. This tip can be moved with an accuracy of $10$~nm and the systematic mapping of the change of the macroscopic electron flow through the device due to the position of the tip allows one to visualize the transport pathways. The experimental data of Aidala et al.\ for a device of area $4 \mu$m$^2$ is shown in Ref.~\cite{Aidala2007a}, together with the theoretical simulations done by wave-packet runs, described in more detail in Ref.~\cite{Kramer2008a}.

\subsection{Aharonov-Bohm interferometer}

As the second application of the wave-packet method in mesoscopic nanodevices, we consider an electron-interferometer in the presence of a magnetic field. The Aharonov-Bohm effect \cite{Aharonov1959} allows one to detect the interference of coherent electrons in a two-path interferometer as oscillations in the probability density. The oscillation-period of the magnetic flux is given by the flux quantum $h/e$. In the nanodevice version discussed here \cite{Webb1985}, the magnetic field also penetrates the arms of the interferometer, and thus the eigenstates do not only acquire a phase, but are also directly distorted by the magnetic field. Using special etching techniques, interferometers with arm-lengths in the $\mu$m-range can be fabricated \cite{Buchholz2010a}.
\begin{figure}[t]
\includegraphics[width=0.7\textwidth]{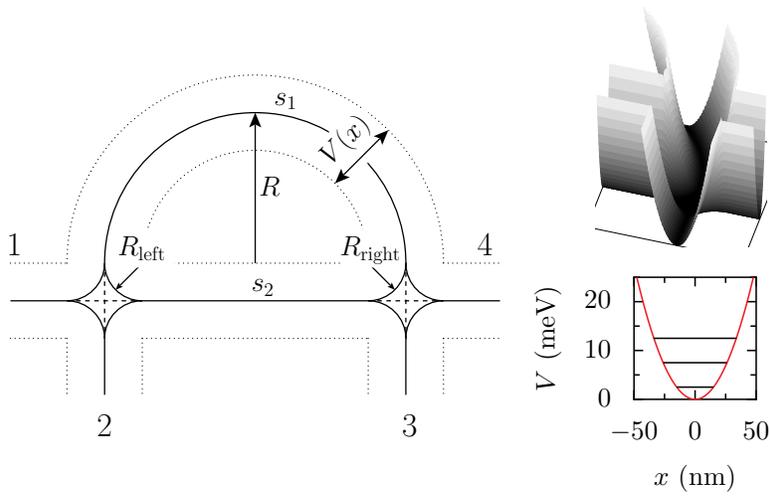}
\caption{Left panel: sketch of the half-circular Aharonov-Bohm interferometer with four attached leads. The right panel shows a close-up of the potential at the crossing, where reflections happen. The steep potential profile along the arms of the interferometer (lower right panel) supports several populated transverse modes below the Fermi energy ($E_F=8-15$~meV). Adapted from Kreisbeck et~al. \cite{Kramer2010b}.\label{fig:abpot}}
\end{figure}
The device in Fig.~\ref{fig:abpot} was designed to facilitate the comparison of theoretical and experimental results and special care was taken to attach to the half-circle leads with minimal imperfections. If the lengths of the two paths $s_1$ and $s_2$ differ, an additional wave-number-dependent phase occurs, given by $\Delta\alpha=k_\mathrm{F}(s_2-s_1)$, where $k_F=\sqrt{2mE_F}/\hbar$ denotes the Fermi wave-number. Ideally, the transmission probability along the paths becomes $T\propto \mathrm{cos}(e\phi/\hbar+\Delta\alpha)$, with magnetic flux $\phi={\rm area}\cdot{\cal B}$ through the enclosed area. The wave-number can be controlled by a perpendicular electric field applied via a top-gate electrode. The simple linear relation between wave-number and phase $\Delta\alpha$ does not take into account time-reversal symmetry, which enforces $T(\MGF)=T(-\MGF)$ in rings with two connecting leads \cite{Onsager1931,Casimir1945,Buettiker1988b} and thus no continuous phase shifts can be achieved. In order to break the phase rigidity, it is necessary to reduce the device symmetry by attaching additional leads to the ring \cite{Buettiker1988b}. The addition of leads increases scattering effects in the cross-junctions and requires to model the device in a two-dimensional fashion. The effective two-dimensional potential at the crossings is sketched in the right panel of Fig.~\ref{fig:abpot}.

Several recently developed recursive Green's function methods principally allow one to compute the transmission through Aharonov-Bohm rings \cite{Kazymyrenko2008a,Wurm2010a} but yield the transmission matrices only for a single Fermi energy. Time-dependent methods based on wave-packet dynamics have been implemented for ring structures \cite{Chaves2009a, Szafran2005a}, but have the disadvantage that merely the transmission of a certain pulse is detected. The time-dependent approach presented in Ref.~\cite{Kramer2010b} allows us to obtain the transmission for a whole range of energies and is thus ideally suited to describe experiments scanning a wide-range of parameters.
\begin{figure}[t]
\includegraphics[width=0.98\textwidth]{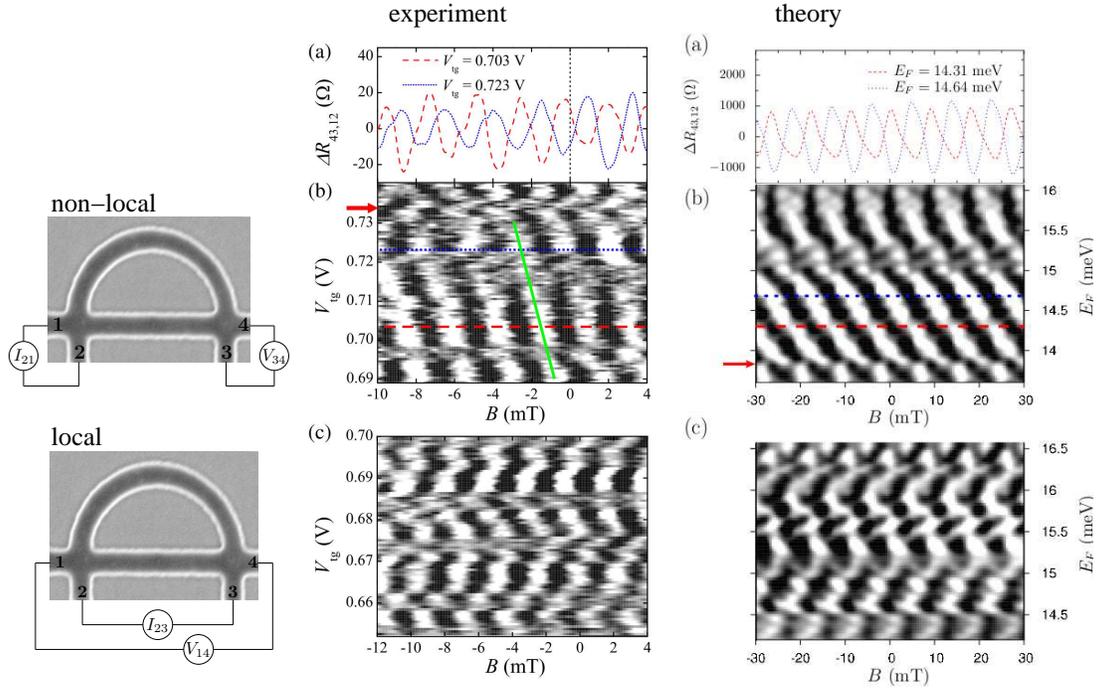}
\caption{Aharonov-Bohm interferometer and interference pattern, comparison of experimental measurements (middle panels) and theoretical calculation (right panels) for two-different voltage and current probe arrangements (upper panels: non-local setup, lower panels local setup). The gray-scale denotes the resistivity of the device. Black and white correspond to maxima and minima of the Aharonov-Bohm transmission amplitudes and the evolution of the extremal value tracks the phase. Sudden phase jumps are indicated by small arrows. The theoretical calculation are performed for a smaller device resulting in a larger magnetic field period. The general features of the experiment and the sudden phase-jumps are reproduced by the theoretical simulations and are related to multiple reflections within the Aharonov-Bohm device. Adapted from Kreisbeck, Kramer (theory) and Buchholz, Fischer, Kunze (experiment) \cite{Kramer2010b}.\label{fig:abosc}}
\end{figure}
The theoretical frame-work for the calculation of the macroscopic voltages and currents through the device is the multi-terminal Landauer-B{\"u}ttiker formalism \cite{Buettiker1988a}, which leads to the following expression for the current from channel $i$
\begin{equation}\label{eq:current}
I_i=\frac{e}{h}\int_{-\infty}^{\infty}\,\rmd E\ \sum_{j \neq i, n_i, n_j}
|t_{in_i\,jn_j}(E)|^2\left( f(E,\mu_i,T)-f(E,\mu_j,T)\right),
\end{equation}
where  $t_{in_i\,jn_j}$ denotes the transmission amplitude for scattering from the transverse mode $n_j$ in arm $j$ into the mode $n_i$ in arm $i$. The Fermi functions $f(E,\mu,T)=(e^{(E-\mu)/k_BT}+1)^{-1}$ characterize the macroscopic reservoirs at the contacts. The semi-infinite lead-channels are assumed to be free of imperfections which cause scattering and allow for constructing well-defined asymptotic channel eigen-state. A similar concept stands behind Wigner's and Eisenbud's R-matrix approach, where an artificial boundary is introduced at the interface between the asymptotic regions and the interaction/scattering region. Within the system the probability current is conserved and all equations can be derived by considering the probability flux through a closed surface. The R-matrix approach describes stationary \cite{Moshinsky1951a} as well as time-dependent processes \cite{Moshinsky1951b}. The channel-eigenstates extend in principle along the semi-infinite lead. For numerical applications, we construct a wave packet of finite extent by forming a superposition of plane waves along the waveguide with a specific transversal mode of the waveguide \cite{Kramer2010a}. Two different measurement setups can be used to probe the oscillatory behavior of the transmission.
In the local-setup (lower panels in Fig.~\ref{fig:abosc}) the time-reversal symmetry is only partially broken by the finite voltages and the unavoidable device-imperfections, whereas in the non-local setup it is possible to adjust the phase by changing the wave-number. The middle panels depict experimental results by Buchholz et al.\ and show that in all cases sudden phase-jumps in the signal occur. Our theoretical calculations (right panels) show a similar behavior. The phase-jumps can be traced back to multiple reflections along the crossings and a resonant coupling of the eigenmodes of the wave-guide and resonant states in the crossings \cite{Kramer2010b}.

\section{Interacting many-body systems: dynamical point of view}

\begin{figure}[t]
\includegraphics[width=0.75\textwidth]{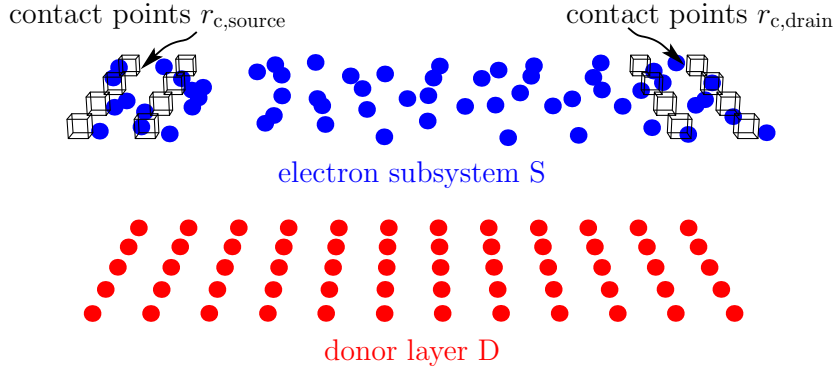}
\caption{\label{fig:devicegeom}Schematic sketch of the device. The potential is tracked at the contact observations points, where particles are injected and removed in accordance with the potential boundary-conditions.}
\end{figure}

Two aspects complicate calculations for interacting many-body systems: the requirement of (anti-)symmetrization of the total wavefunction including spin, and the inclusion of the Coulomb forces. Exact results for interacting quantum-systems are difficult to obtain. The case of two-interacting electrons provides already interesting insights into antisymmetrization effects and is discussed in detail in Ref.~\cite{Kramer2010c} using different analytical methods, and also numerical and variational approaches. For a larger particle numbers (>100), a complete quantum-mechanical description exceeds available computer power. Transport calculations face additional difficulties, since in principle the complete electrical circuit (including the power source) has to be considered. In practice, at a certain point we have to make a cut between the microscopic and the macroscopic description. The formulation of the correct boundary conditions at this cut presents a formidable challenge and an unsolved problem. Most approaches introduce an asymptotic region, where the Coulomb-forces are gradually switched off, but it is not clear if this procedure captures the nature of a real physical device. 
In this section we discuss an intriguing problem of condensed matter theory, the classical Hall effect \cite{Hall1879a,Kramer2009c}. The formulation of realistic boundary conditions is already required in a classical many-body theory. The computational power for solving classical many-body problems has jumped to new levels over the last years due to the availability of general purpose \textbf{G}raphics \textbf{P}rocessing \textbf{U}nits (GPU). Driven by the ever increasing demand for realistic rendering of computer-graphics in games, the processor development of the GPU has overtaken speedwise the standard CPUs. Conceptionally, GPUs consists of several hundred stream processing units (on an NVIDIA C2050 board 448 units), which work in parallel. The total number of floating point operations per second exceeds $1.2\times10^{12}$. GPU programs can be written in the Open Compute Language (OpenCL), which allows to use a C++ syntax for the main-program, while the GPU routines are coded in a special language. For the Coulombic problem at hand the fast build-in reciprocal square-root operation of the GPU yields another significant speed gain.

\begin{figure}[t]
\includegraphics[width=0.65\textwidth]{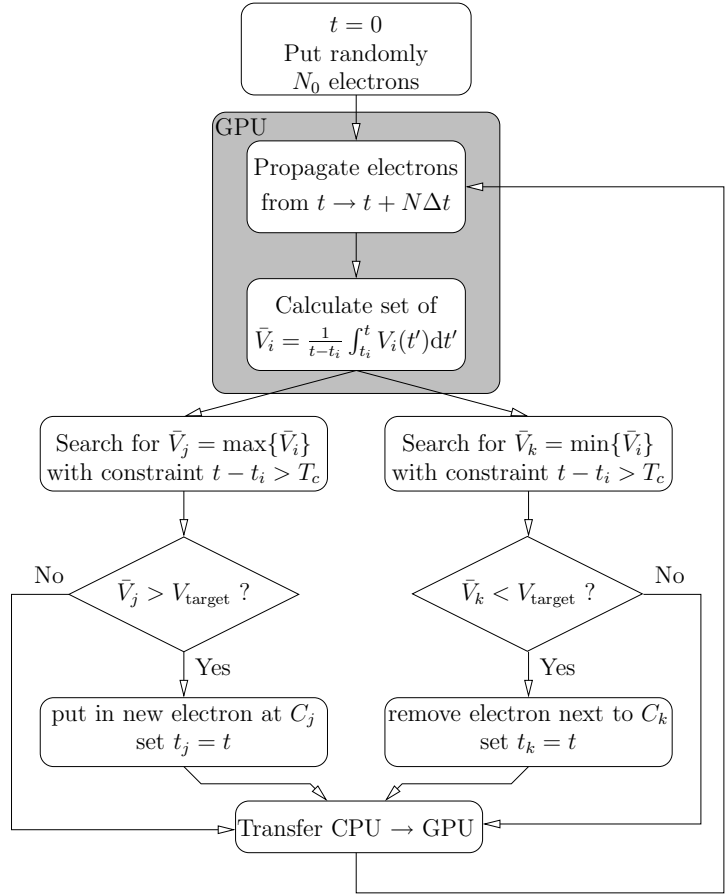}
\caption{\label{fig:progflow}Program flow to obtain the self-consisten Hall potential from a numerical N-body simulation using a hybrid GPU-CPU scheme. The injection/removal events are timed and spatially distributed to maintain an equipotential surface at the contacts. Adapted from Ref.~\cite{Kramer2009c}.}
\end{figure}

\subsection{Modelling a Hall device}

Hall nanodevice can be fabricated from AlGaAs/GaAs heterostructures, discussed in the previous section. The Hall-bar is schematically sketched in Fig.~\ref{fig:devicegeom} and consists of several interconnected regions. The positively charged donor-layer is shown at the bottom and consists of 8094 charges. At a vertical distance of $10$~nm the electron layer is situated, where electrons can move in a two-dimensional rectangular area of length $2.5$~$\mu$m and width $1.0$~$\mu$m. At the left-end (source) and right-end (drain) of the electron layer, potential observation points are placed. The observation points track the time-evolution of the potential due to all the fixed and moveable charges in the device. Depending on the observed potentials, electrons are injected or removed.
\begin{figure}[t]
\includegraphics[width=0.7\textwidth]{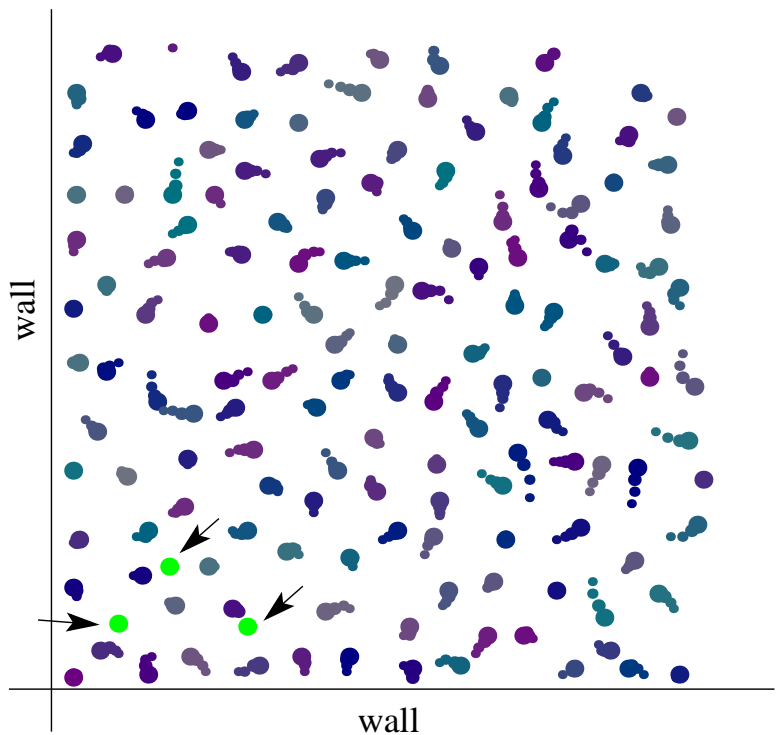}
\caption{\label{fig:snapshot}Snapshot of the electron distribution near the source contact. Arrows mark recently injected electrons, which are required to maintain the equipotential within the contact region.}
\end{figure}
The program flow is shown in Fig.~\ref{fig:progflow} and consists of the following steps:
Initially we populate the device with randomly distributed electrons. 
Next we calculate the forces acting on each electron and update the electron positions by one time-step. The force $F_k$ on the $k$th electron consists of the Coulombic forces $F^{\rm C}_k$ due to all other $N_e$ electrons, $N_d$ donors, and the velocity dependent Lorentz force $F^{\rm L}_k$,
\begin{eqnarray}\label{eq:forces}
\mathbf{F}_{k}^C&=&
-\frac{q^2}{4\pi\epsilon_0 \epsilon} \sum_{\substack{l=1\\l\ne k}}^{N_e}
 \frac{\mathbf{r}_l-\mathbf{r}_k}{|\mathbf{r}_l-\mathbf{r}_k|^3}
+\frac{q^2}{4\pi\epsilon_0 \epsilon} \sum_{l=1}^{N_d}
 \frac{\mathbf{r}_l-\mathbf{r}_k}{|\mathbf{r}_l-\mathbf{r}_k|^3}
\\
\mathbf{F}_{k}^L&=&q\;\dot{\mathbf{r}}_k\times\MGF,
\end{eqnarray}
where $\mgf$ denotes the magnetic field, pointing perpendicular to the device layer. Each electron state-vector contains the position $\mathbf{r}_k$, and the velocity $\dot{\mathbf{r}}_k$. The equation of motions are integrated with Boris' algorithm \cite{Verboncoeur2005},
\begin{subequations}
\begin{align}
\dot{\mathbf{r}}_k(t + \Delta t) =& 
\frac{1-\omega_l^2 \Delta t^2}{1+\omega_l^2 \Delta t^2} \dot{\mathbf{r}}(t) +
\frac{2 \omega_l \Delta t}{1+\omega_l^2 \Delta t^2} 
\begin{pmatrix}
0 & -1 \\
1 & 0
\end{pmatrix}
\dot{\mathbf{r}}(t)
+\frac{\mathbf{F}_k^{\rm C}(\mathbf{r}_k(t))}{m_e}\\
\mathbf{r}_k(t+\Delta t) =& 
\mathbf{r}_k(t) + \Delta t~\dot{\mathbf{r}}_k(t + \Delta t),
\end{align}
\end{subequations}
where $\omega_l = \frac{e\mgf}{2 m_e}$ denotes the Larmor frequency. Typical time-steps are $1/5000$ of the cyclotron period, which amounts to $\Delta t=5\times 10^{-17}$~s for a magnetic field of ${\cal B}=4$~T and an effective electron mass $m=0.067 m_e$.
\begin{figure}[t]
\includegraphics[width=0.7\textwidth]{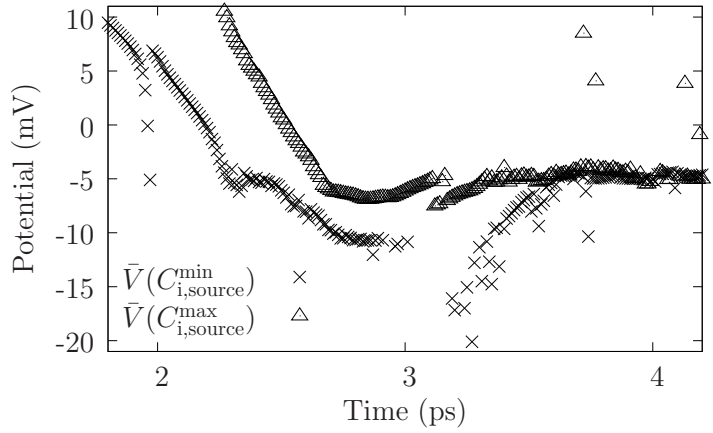}
\caption{\label{fig:potcon}Extremal values of the potential in the source contact, which converges towards the target value of $V_{\rm source}=-5$~mV.}
\end{figure}
The observation points at the source and drain contacts $\mathbf{r}_{{\rm source},i}$, $\mathbf{r}_{{\rm drain},i}$ record the local potential averaged over a time-period $\Delta t_{\rm av}$
\begin{equation}
\overline{V}(\mathbf{r}_i)=\frac{1}{4\pi\epsilon_0 \epsilon \Delta t_{\rm av}}
\int_{t-\Delta t_{\rm av}}^{t}\rmd t'\; \sum_{\substack{l=1\\l\ne i}}^N
 \frac{q_l}{|\mathbf{r}_l(t')-\mathbf{r}_i|},
\end{equation}
where the sum runs over all electrons and donors present in the system. The contacts serve as interfaces to electron reservoirs and operate under the condition that an equipotential is required within the contact-pad. The equipotential condition is physically motivated by the model of metallic contacts. In practice, we establish an (on average) constant potential by injection an electron at the observation point, which shows the biggest positive deviation from the requested contact potential. Similarly, an electron is removed at the observation point which shows the biggest negative deviation from the contact potential. If none of the conditions is met, the electron propagation continues without injection or removal events.
\begin{figure}[t]
\begin{minipage}{\textwidth}
\begin{center}
\includegraphics[width=0.7\textwidth]{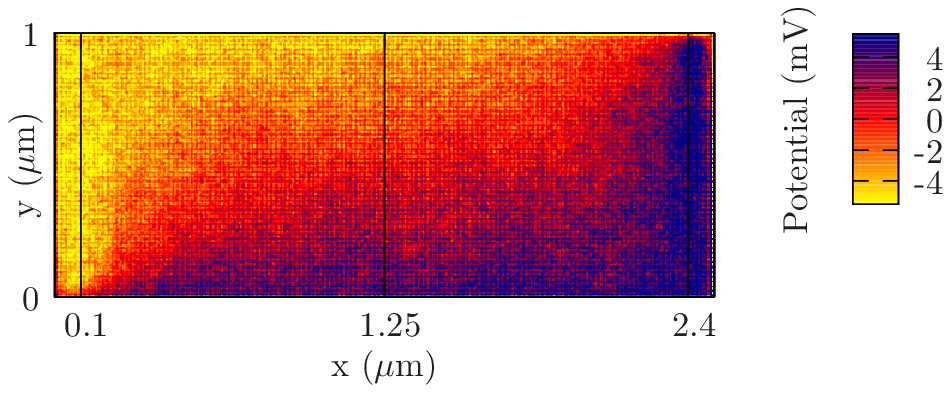}\\[3ex]
\includegraphics[width=0.6\textwidth]{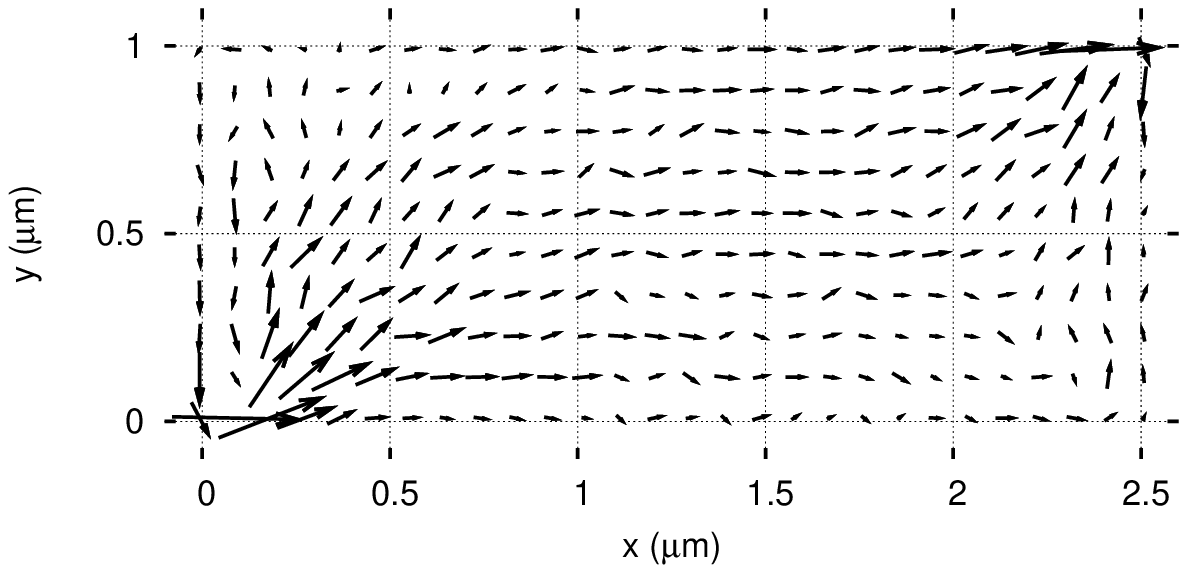}
\end{center}
\end{minipage}
\caption{\label{fig:potav}Upper panel: Time-averaged Hall potential between the source (left end) and the drain (right end) contact. Lower panel: current density distribution showing the transport pathway and the emergence of two ``hot-spots'' at opposite corners of the device. In the left hot-spot electrons are injected into the device.}
\end{figure}
The extremal deviations from the prescribed source-contact (here $V_{\rm source}=-5$~mV) value are shown in Fig.~\ref{fig:potcon}. Convergence is reached after $0.04$~ns and the complete system approaches a steady-state, where the total injection rate and the total removal rate approaches constant values. During the simulation, the state-vector of the complete system is stored periodically each 10000 integration steps. After the simulation run, a statistical analysis is performed on a regular grid covering the complete device area. At each grid point the time-averaged potential $V_{av}$ is obtained, typical averaging times are in the range of $T_{\rm av}=1$~ns, corresponding to 1000 stored state-vectors of the system. The time-averaged potential is shown in Fig.~\ref{fig:potav}. The long-time average reveals an S-shaped potential, which emerges from in a self-consistent way from the many-body calculation. The shape of the potential is a direct consequence of the metallic boundary conditions, the particle-interactions, and the specific device geometry. Experimentally, the same potential has also been observed in the quantum Hall effect (QHE) \cite{Knott1995a,Ahlswede2001a}. If we take the observed and theoretically calculated potentials as the mean-field potential of the device, we can quantize the system and obtain a theory which contains the integer quantum Hall effect \cite{Kramer2009b}. The electric field is strongest in the vicinity of the current source contact and leads to a broadening of the local density of states \cite{Kramer2003a}. In the time-dependent picture such a broadening signifies a decaying autocorrelation function $C(t)\rightarrow 0$ and thus an extended state which guides the electron through the complete device to the drain contact.

\section{Conclusion}

The time-dependent approach to complex systems provides an intuitive and dynamical picture of transport phenomena. 
The replace stationary state problems by propagation from initial states, in this way we avoid computationally expensive matrix-diagonalization methods, and by the wave-packet method provide results for a wide range of energies with a single propagation. For interacting many-body systems, graphics processing units are powerful tools to devise microscopic models and to test the validity of macroscopic equations.

\subsection*{Acknowledgments}

I thank the organizers of the Latinamerican School of Physics for the invitation to present this course in Mexico City and for their kind hospitality. The course and this contribution is dedicated to the memory of Marcos Moshinsky. Financial support from the Emmy-Noether program of the Deutsche Forschungsgemeinschaft (grant KR 2889/2) is gratefully acknowledged. I appreciate helpful discussions with my collaborators, Manfred Kleber, Eric J.\ Heller, Christian Bracher, Robert E.\ Parrott, Christoph Kreisbeck, and Viktor Krueckl. Many graphs and papers are the outcome of the collaborations and discussions with experimentally working colleagues, in particular Christophe Blondel, John Yukich, Katherine Aidala, Robert Westervelt, Sven Buchholz, Saskia Fischer, and Ulrich Kunze.

\bibliographystyle{aipproc}
\providecommand{\url}[1]{#1}

\end{document}